  \documentclass[mnsc,nonblindrev]{informs3} 

\OneAndAHalfSpacedXII 

\JOURNAL{}

\usepackage{natbib}
 \bibpunct[, ]{(}{)}{,}{a}{}{,}%
 \def\bibfont{\small}%
\TheoremsNumberedThrough     
\ECRepeatTheorems

\EquationsNumberedThrough    

\MANUSCRIPTNO{}

\usepackage{hyperref}
\usepackage{paralist}
\usepackage{xspace}
\usepackage{tabularx}
\usepackage{booktabs}
\usepackage{graphicx}
\usepackage{xcolor}
\usepackage{subfig}
\usepackage{enumitem}
\usepackage{float}
\usepackage{rotating}
\usepackage{bbding} 

\newcommand{\hide}[1]{}

\newcommand{\bit}{\begin{compactitem}}
\newcommand{\eit}{\end{compactitem}}
\newcommand{\ben}{\begin{compactenum}}
\newcommand{\een}{\end{compactenum}}

\newcommand{\RNum}[1]{\uppercase\expandafter{\romannumeral #1\relax}}

\newcommand{\medicare}{Medicare\xspace}
\newcommand{\pps}{PPS\xspace}
\newcommand{\doj}{DOJ\xspace}
\newcommand{\icd}{ICD\xspace}
\newcommand{\drg}{DRG\xspace}
\newcommand{\mdc}{MDC\xspace}

\newcommand{\bs}[1]{\boldsymbol{#1}}

\newcommand{\x}{\bs{X}\xspace} 
\newcommand{\y}{Y\xspace}
\newcommand{\provider}{H\xspace}
\newcommand{\nprovider}{N_H\xspace} 
\newcommand{\nicd}{M_H\xspace} 
\newcommand{\pvec}{\bs{v}\xspace} 
\newcommand{\qvec}{\bs{q}\xspace} 
\newcommand{\cost}{Cost\xspace} 
\newcommand{\excess}{ExcessSpending\xspace} 

\newcommand{\ifr}{\textsc{iF}\xspace}
\newcommand{\sod}{\textsc{SOD}\xspace}
\newcommand{\rshash}{\textsc{RSHash}\xspace}
\newcommand{\loda}{\textsc{LODA}\xspace}
\newcommand{\rrcf}{\textsc{RRCF}\xspace}

\newcommand{\isout}{\textsuperscript{*}\xspace}

\newcommand{\acktext}{Research reported in this publication was supported by the National Institute on Aging of the National Institutes of Health under Award Number P30AG012810. The content is solely the responsibility of the authors and does not necessarily represent the official views of the National Institutes of Health. We thank the NBER for data access and support and Lowell Taylor for his contributions to earlier stages of the project.\xspace}

\usepackage{xcolor}

\begin{document}


 \RUNAUTHOR{Shekhar, Leder-Luis, Akoglu}

 \RUNTITLE{Unsupervised ML for Explainable Health Care Fraud Detection}
\TITLE{Unsupervised Machine Learning for Explainable Health Care Fraud Detection\footnote{\acktext}\\%
}

\ARTICLEAUTHORS{%
\AUTHOR{Shubhranshu Shekhar}
\AFF{Carnegie Mellon University, \EMAIL{shubhras@andrew.cmu.edu}} 
\AUTHOR{Jetson Leder-Luis}
\AFF{Boston University and NBER, \EMAIL{jetson@bu.edu}}

\AUTHOR{Leman Akoglu}
\AFF{Carnegie Mellon University, \EMAIL{lakoglu@andrew.cmu.edu}}

} 

\ABSTRACT{%
The US spends more than 4 trillion dollars per year on health care, largely conducted by private providers and reimbursed by insurers. A major concern in this system is overbilling, waste and fraud by providers, who face incentives to misreport on their claims in order to receive higher payments. In this work, we develop novel machine learning tools to identify providers that overbill insurers. Using large-scale claims data from Medicare, the US federal health insurance program for elderly adults and the disabled, we identify patterns consistent with fraud or overbilling among inpatient hospitalizations. Our proposed approach for fraud detection is fully unsupervised, not relying on any labeled training data, and is explainable to end users, providing reasoning and interpretable insights into the potentially suspicious behavior of the flagged providers. Data from the Department of Justice on providers facing anti-fraud lawsuits and case studies of suspicious providers validate our approach and findings. We also perform a post-analysis to understand hospital characteristics, those not used for detection but associate with a high suspiciousness score. Our method provides an 8-fold lift over random targeting, and can be used to guide investigations and auditing of suspicious providers for both public and private health insurance systems.
}%



\maketitle
\newpage

%


\section{Introduction}
\label{sec:intro}
Fraud in health care is hard to detect. Insurers face information asymmetries, where physicians and patients both know more about the health care delivered than the insurer responsible for paying for that care. Health care providers such as doctors and hospitals face incentives to maximize their reimbursements from health insurance companies, and insurers must largely rely on documentation from providers themselves. This asymmetric information leads to circumstances where unscrupulous providers can choose to commit fraud.\looseness=-1

 The scale of health care spending means that even small amounts of fraud can be very expensive. Estimated US health care spending in 2019 was \$3.81 Trillion~\citep{nhe2021fact},
almost as high as the GDP of Germany, the 4th largest in the world.  
National health care spending in the US is  expected to grow at an average annual rate of 5.4\%\footnote{The growth was a striking 36.0\% in 2020 in response to the COVID-19 pandemic.}, from 2019 to 2028, outpacing US GDP at 4.3\%. Efforts to detect and root out fraud are paramount for limiting the growth of wasteful spending.

These issues are compounded in the federal health care programs, where the government is the insurer. The US federal government spends over a trillion dollars per year on health insurance, largely paid to private firms, and fraud detection is challenging due to the sheer volume of claims being processed. The largest of these programs is Medicare, the federal  health insurance program for people of age $65$ and older and the disabled. With more than \$800 Billion spent on Medicare in 2019, even small shares of waste and abuse lead to large losses, which are ultimately paid for by taxpayers and reduce the capacity of the government to fund valuable social programs.   
The US Government Accountability Office (GAO) estimates Medicare improper payments, a measure of mistaken or inappropriately documented spending, in 2019 at \$46.2 Billion~\citep{gao2019payment}.
This problem has gained the attention of Medicare administrators faced with the challenge of detecting and deterring waste and fraud to ensure the program stays financially solvent~\citep{health2022annual}. 

The nature of health care fraud provides insights into how it can be detected. Healthcare providers face incentives to manipulate billing to increase profits.  Yet, in general, patients see multiple providers, and there are many providers in the system that do not commit fraud. Therefore, fraud detection does not rely on the verification of any particular claim, but rather detecting provider-level patterns of care that appear anomalous when considering patient characteristics, medical history, and patterns of behavior by regular non-fraudulent providers.

In this work, we develop new tools to detect health care overbilling or fraud.  We build a machine learning (ML) framework to discover patterns and detect anomalous providers using large-scale Medicare claims data.  Our method focuses on inpatient hospitalization, the largest category of spending and the highest-intensity health care provided by Medicare, which cost the US government \$147 Billion in 2021. The proposed approach identifies anomalous providers based on their billing patterns, using patient-level data including medical history, demographics, and geography. We employ our method to identify anomalous patterns among providers and rank them in order of their suspiciousness \textit{without using any supervision}, that is, not relying on any {\it a priori} labeled training data. Moreover, our approach is equipped with \textit{explanations} to the suspiciousness of the flagged providers, enabling end users like auditors to use our results to guide further investigation.

Our approach is an ensemble method, utilizing three novel unsupervised detection algorithms that uncover aberrant patterns in care across different data modalities. The first component of the ensemble focuses on providers\footnote{In this work, the words provider and hospital are used interchangeably. While providers can refer to any health care service provider, we specifically study hospitals.} with large observed expenditures conditioned on patient characteristics and medical history. We use a regression-based analysis to identify providers with large fixed effects that correspond to high spending per patient even controlling for observable medical history and location of the patient. The second component focuses on  coding behavior of claims, uncovering rare ICD-10 medical coding patterns employed by providers, which is indicative of manipulation of specific codes a patient is tagged with in order to garner higher reimbursements. The third component is peer based, focusing on identifying aberrant hospital billing code (DRG) patterns among a related group of hospitals, where the group of hospitals share similar patient populations and distributions of types of care. 

We assemble the evidence from these  
three detection methods together to rank providers based on suspiciousness. We utilize instant-runoff voting~\citep{franceschini2022ranking} to reach an aggregate ranking for the suspiciousness of providers. This method follows an iterative procedure to rank the hospital that is most suspicious based on the ``vote'' across different detectors in each round. 

We validate our approach quantitatively with ground-truth data from the Department of Justice (DOJ). Using a corpus of thousands of DOJ press releases about fraud, we tag providers identified as fraudulent and merge these data to compare with our ranking. While only 1 in 20 hospitals nationwide are named in the DOJ Press releases, our ranking substantially improves detection over random sampling: the top 50 providers identified by our method contain 21 providers named in the same DOJ corpus, that an 8-fold lift in detection rate.  We note that providers ranked high by our method but not listed by the DOJ are not necessarily false positives; rather, enforcement by the DOJ reflects a combination of opportunity to enforce and capacity constraints, and hence only provides partial ground-truth. The DOJ validation is a form of positive-unlabeled data \citep{bekker2020learning}, and the overlap with our method is therefore a lower-bound of the amount of fraud successfully detected.\looseness=-1

In summary, our proposed approach provides scalable and explainable tools to detecting fraud and abuse in health care systems. Our method does not rely on any supervision or data labeling labor, and thus can be readily employed on massive unlabeled data. As the detectors utilize different data modalities and modeling approaches, the explanations also provide different perspective and reasoning into suspicious behavior. This makes our proposed approach useful in practice, as auditors would be presented with multiple pieces of evidence that support a case and can aid with further investigation. While our analysis focuses on hospitals, this method could be readily adapted for use in detecting overbilling in outpatient claims, doctor's office visits, or other areas of potentially fraudulent care.

We foresee that our method could be particularly effective at auditing of health care providers and guiding future enforcement. While the data set on which we build our method is from Medicare, we anticipate our methods will prove valuable to private insurers as well, who face nearly identical challenges in eliminating fraud from private health insurance systems. As our ranking provides a significant lift in detection rate than one would achieve by random sampling, it can be used to target and prioritize auditing. While our explanations cannot provide legal-standard evidence of bad behavior by providers, they can help sense-making and be used as starting points that guide deeper investigation. Overall, we anticipate that our proposed solution will have value for policymakers, auditors, and enforcers in the health care domain at large. 

This paper proceeds as follows. We describe the background and institutions regarding Medicare payment, health care fraud, and fraud enforcement in Section~\ref{sec:background}, followed by a description of Medicare data in  Section~\ref{sec:data}. Then, we present our detection and explanation methodologies, with an overview in Section \ref{sec:meth}. 
Section~\ref{sec:regression} presents the global expenditure regression-based OD model; Section~\ref{sec:subspace} presents the local ICD subspace based OD model; and Section~\ref{sec:peers} presents the local/contextual peer-based excess cost OD model.  Section \ref{sec:aggregate} reports the ensemble model detection results and multi-view explanations on several case studies. 
Finally, Section \ref{sec:desc} provides a post-analysis toward characterizing hospitals with high estimated suspiciousness.
We conclude with discussion and takeaways in Section \ref{sec:concl}.



\section{Background}
\label{sec:background}

In this section, we discuss the institutional details of Medicare fraud.  First, we describe the Medicare payment system for inpatient hospitalization, which creates incentives for fraud. Second, we discuss the various types of Medicare fraud and the ways in which it is enforced. While many of the institutional details about Medicare claims and enforcement are specific to the federal system, the general nature of health care billing is consistent across both publicly funded and private-payer systems.

\subsection{Medicare Payment System}

Medicare uses a prospective payment system (\pps) for inpatient hospitalization, where providers are paid a fixed amount for each patient's stay, regardless of stay length or cost. Patients are coded with diagnoses and procedure codes based on the International Classification of Diseases (ICD) system, and then based on this coding, each inpatient stay is classified into one Medicare Severity Diagnosis Related Group (DRG). Each DRG is associated with a certain fixed amount per stay, with possible small adjustments \citep{medpac2021hospital}. The fixed payment for each DRG is based on the average costs of treating patients under that DRG code nationwide and it is updated annually.

The \pps incentivizes providers to keep the healthcare costs down~\citep{ellis1986provider} since the provider's profit is the difference between the fixed DRG payment and the treatment cost.  This is in contrast to a reimbursement-based system, where providers would face incentive to incur higher costs for higher reimbursement. However, the \pps may lead to hospitals trying to avoid treating high-cost patients. To address such issues, \pps adjusts the DRG payment~\citep{medpac2021hospital} to include provider specific factors such as provider's wage index (geographic factor), patient case-mix to account for patient-population specific treatment cost, teaching and research expenditure, disproportionate share of low-income patients, and number of unusually costly outlier cases. 

\subsection{Health Care Fraud}

Hospitals face incentives to miscode patients; when done intentionally or recklessly, this can qualify as fraud.  Because the patient's ICD coding dictates their DRG and ultimately the hospital reimbursement amount, hospital coding decisions directly affect hospital profits. 

Fraud in inpatient hospitalization takes many forms.  One well-studied form is \textit{upcoding}, where hospitals miscode patients to higher severity levels of care in order to receive higher reimbursement~\citep{dafny2005hospitals,silverman2004medicare,becker2005detecting}. A second common issue is \textit{lack of medical necessity}, where a patient's health conditions do not qualify them for that care \citep{howard2020false}. Moreover, there is a variety of conduct that can also qualify as health care fraud, such as providing compensation to providers for referring patients, which qualifies as a {\it kickback}. 

In this paper, we are largely agnostic to which type of fraud hospitals commit, and instead focus on payment levels. In general, fraud is of greatest concern when it results in wasteful spending. Our method detects hospitals whose anomalous conduct results in higher payments, which is valuable for detecting hospitals where additional auditing is of highest marginal value.

\subsection{Health Care Anti-Fraud Enforcement}
\label{ssec:antifraud}

The US government undertakes a number of initiatives to detect and deter waste, fraud and abuse in federally-funded health care spending.  Our method, which relies solely on claims data, is complementary to existing methodologies. Private insurers face similar challenges and also work to detect, investigate and enforce against fraudulent providers, although they lack the full weight of the federal investigatory system. 

Federal law prohibits Medicare fraud and provides avenues by which fraud can be addressed through criminal and civil enforcement.  The federal health care fraud statute provides criminal penalties for those who commit health care fraud, and this enforcement is compounded by criminal enforcement under the anti-kickback statute, as well as the wire fraud and racketeering statutes.  Criminal Medicare fraud is prosecuted by the Department of Justice. For a deeper treatment of criminal Medicare fraud, see~\cite{eliason2021ambulance}.

Civil enforcement for Medicare fraud operates through the False Claims Act, which provides an avenue for whistleblowers to come forward with information about fraud and receive compensation.  Whistleblowers file their own cases in federal civil court, and the DOJ has an option to support these cases. 
\cite{leder2020can} and
\cite{howard2020false}
provide more information about the False Claims Act and show that these whistleblowers provide high deterrence effects.

In addition to litigation, administrators use a variety of policy tools to limit health care waste, fraud and abuse. The Office of the Inspector General of Health and Human Services undertakes administrative actions against firms that overbill Medicare. Medicare also has a variety of auditing programs that seek to detect unnecessary or unjustified spending; see~\cite{shi2022monitoring} for a description of the Recovery Audit Contractors program.  Finally, Medicare uses regulations to target unnecessary spending, such as prior authorization requirements. Some of these regulations combat fraud while others combat waste; see~\cite{brot2022rationing} and \cite{eliason2021ambulance} for a discussion of these regulations.  

In addition to the enforcement actions listed above, Medicare and private insurers undertake some data-driven investigatory work in order to detect fraud. These efforts have received little attention in academic work. Medicare claims processors work with contractors called Unified Program Integrity Coordinators (UPICs)~\citep{noridian2022unified} to audit and detect aberrant payments. In addition, Medicare uses a private-public partnership model through the Healthcare Fraud Prevention Partnership to share data between the federal government and private insurers to detect health care fraud with patterns similar across a variety of types of care and different health insurance programs~\citep{partnership2022healthcare}.  When fraud is identified through these data-driven efforts, investigators can refer those cases to the DOJ for civil or criminal prosecution.\looseness=-1

In this paper, we curate a list of hospitals that have been subject to DOJ actions at both the criminal and civil level, used for quantitative evaluation of our method. While there are many ways in which hospitals could have been investigated or sanctioned, being named in a DOJ press release validates that the hospital was likely committing behavior that rose to the level of criminal or civil fraud, which represents a true positive. A disclaimer, on the other hand, is that the  hospitals subjected to DOJ actions likely constitute only a partial list of all fraudulent hospitals, as other unknown fraud and waste may have gone undetected, which represents a false negative.

\subsection{Related Methodological Work}


In addition to the economic studies listed above that discuss health care fraud, several data-centric approaches have been explored in the context of Medicare fraud. We refer the reader to~\cite{bauder2017survey,kumaraswamy2022healthcare,joudaki2015using} for detailed survey on different methods. 

In early work, \cite{rosenberg2000statistical} study upcoding within the claims data. They estimate the probability that a claim has incorrect DRG code, which they further use to identify claims to investigate and audit. \cite{brunt2011cpt} study upcoding in the physician office visits data, where they estimate the likelihood of a disease code selected for an office visit. They study the payment differential in the selected code and code used in the data to understand the upcoding practices. \cite{fang2017detecting} find evidence of provider overbilling using inappropriately high number of hours worked to identify outliers.

Recently, \cite{chandola2013knowledge,suresh2014detection} introduce methods for provider profile comparison to spot possible misuses or fraud. These works focus on introducing methods and features to represent hospital profiles for comparison, however, do not present any conclusive results. On the other hand, \cite{bauder2018medicare,bauder2018detection,herland2018big,bauder2017medicare} utilize publicly available excluded providers to learn models for detection of fraudulent providers. However these approaches rely on availability of human labeled information on fraudulent information, which is often incomplete and hard to obtain for massive Medicare data. 

In contrast to earlier methods, unsupervised and explainable methods for the problem, which are more practical in the real world, have received limited attention. \cite{luo2010unsupervised} compare DRG distributions of hospitals providing services for hip replacements and heart health to differences in coding. The underlying assumption is that most hospitals will have similar distribution conditioned on the treatment provided. Recently, \cite{ekin2019unsupervised} learn joint distribution of medical procedures and providers using outpatient data. The joint distribution is used to identify provider anomalies based on procedure code and usage frequency by the provider.
Most of the work uses only a fraction of massive Medicare data, and often do not incorporate an explanation of results that could be useful to investigators. Our method builds upon these existing studies to provide a precise and explainable detection method that does not rely upon the existence of labeled data.

\section{Data Description}
\label{sec:data}
This study combines data from a variety of sources to detect anomalous provider spending behavior  in Medicare and compare it to ground-truth labeling of providers that have faced anti-fraud enforcement.

Our analysis of provider behavior uses a large-scale dataset of Medicare claims. Data were accessed through a data use agreement with the Centers for Medicare and Medicaid Services, facilitated by the Research Data Assistant Center (ResDAC) and the National Bureau of Economic Research (NBER). These hundreds of millions of observations contain extensive data about each hospitalization and patient in the Medicare system, providing an ideal corpus with which to study hospital behavior.

We consider patients hospitalized in 2017, and we use data from 2012 through 2016 to construct the patients' medical history.  For these years, we use  100\% of samples of Fee-For-Service institutional Medicare data, including inpatient and outpatient claims, and beneficiary \footnote{Patient refers to a person receiving health care; beneficiary refers to a person covered by health insurance. Here, they are used interchangeably, as all of our data come from patients who are Medicare beneficiaries.} information including demographic information and chronic condition indicators from the Chronic Conditions Warehouse. To further understand a beneficiary's medicare history, we use 20\% of  samples of carrier files, which describe physician office visits.\footnote{20\% samples are the largest available for physician office visits.}\looseness=-1

\begin{table}
	\centering
	\caption{Inpatient data statistics from year 2017 \label{tab:datastat}}
	\begin{tabular}{lr}
		\toprule
		\textbf{Spending} & \\
		\midrule
		Medicare total expenditure\citep{statista2022nhetotal} &  \$$710$ billion \\
		Medicare inpatient expenditure &   \$$131$ billion\\
		\midrule
		\textbf{Beneficiaries} &\\
		\midrule
		Number of inpatient beneficiaries & $6.6$ million\\
		Number of inpatient claims & $11.2$ million \\
		\midrule
		\textbf{Providers} & \\
		\midrule
		Number of providers & 7,661\\
		\bottomrule
	\end{tabular}
\end{table}

Table~\ref{tab:datastat} describes the sample of inpatient hospitalization claims from 2017.  We observe 11.2 million claims from 6.6 million beneficiaries representing 7,661 different providers.  Medicare spent in total \$131 billion on inpatient care in 2017, out of \$710 billion total reported Medicare spending. 

Table~\ref{tab:datahistory} describes our sample used to construct patient medical history from 2012 through 2016. We observe nearly a hundred million physician office visits and another hundred million outpatient visits per year, as well as millions of inpatient visits per year. Appendix \ref{appendix:medicaredata} provides additional details about the cleaning and use of the Medicare data. 

To understand provider characteristics, we use the Medicare Provider-of-Service files, which contain details on providers such as certification number, name, the type of Medicare services that it provides, and type of ownership (private or public). We can identify patients across files using their unique beneficiary identifiers, and we identify providers by their identifiers such as the National Provider Identification (NPI) or CMS Certification Number (CCN). Further, we separately identify Academic Medical Centers based on their membership to Council of Teaching Hospitals~\citep{coth2021participants}. 
These providers engage in academic research,  which could lead them to be ranked as anomalous due to the differences in their claim patterns from other hospitals.

\begin{table}
	\centering
	\caption{Scale of data from year 2012 to 2016 used to build medical history of patients who are 70 years or older in the inpatient claims from year 2017. The number in each cell is in millions. \label{tab:datahistory}}
	\begin{tabular}{lrrrrr}
		\toprule
		 & 2012 & 2013 & 2014 & 2015 & 2016\\
		\midrule
		Physician visits & 94.7 & 100.2 & 102.8 & 107.7 & 114.2\\
		Outpatient visits & 81.5 & 87.3 & 90.9 & 96.8 & 104.3\\
		Inpatient visits & 4.0 & 4.2 & 4.4 & 5.1 & 5.8 \\
		\bottomrule
	\end{tabular}
\end{table}



The federal Department of Justice (DOJ) publishes press releases when fraud is identified in order to inform the press and the public as well as deter future fraudulent behavior.
To evaluate our automated detection of suspicious providers, we utilize these press releases related to Medicare from the DOJ. To that end, we scraped from the DOJ website thousands of press releases that contain the word `Medicare'. Each press release corresponds to a case that the Department of Justice was involved with, often at the time of settlement. Using partial name matching, we tag the hospitals that appear in this corpus. As the DOJ lacks both the capacity and the information to prosecute all Medicare fraud, the press releases provide only a partial  list of providers that have engaged in fraudulent behavior. We can consider this a form of positive-unlabeled data: while we can identify firms that have been named in a press release as having likely committed fraud, firms that are {\it not} named are not necessarily above suspicion. Appendix \ref{appendix:dojdata} provides additional details about the collection and cleaning of the DOJ corpus.

\section{Method Overview}
\label{sec:meth}


The Medicare dataset comprises diverse data modalities, which provides an opportunity for modeling the fraud detection problem in various ways. For example, a provider can be represented by the DRG (billing) codes associated with its claims, the frequency of ICD (diagnosis and procedure) codes used in its claims, or by the characteristics of the patient populations that it serves. Each modality presents us with a specific perspective of the data. These different modalities then allow us to learn comprehensive provider behavior which reveal information that cannot be completely uncovered based on only one aspect of the data, since each representation may contain information that is not reflected in others. 

In this work, our goal is to estimate a suspiciousness score based on which we rank providers such that anomalous ones are ranked at the top, which may be due to their fraudulent practices. To utilize our different data modalities, we propose an unsupervised \textit{multi-view} anomaly detection approach, suitable for the underlying multi-modal data. Each view (or base detector) presents itself as a different model of the anomalies, operating on a different data representation. As such, each can be seen as providing evidence that corresponds to a particular reason for detection. The explanation provided by each detector provides a unique perspective into suspicious behavior. Collectively, the evidence from these base detectors, i.e. across modalities, can be assembled systematically into an ensemble detection method. 

\begin{figure}[!t]
    \centering
    \includegraphics[width=\linewidth]{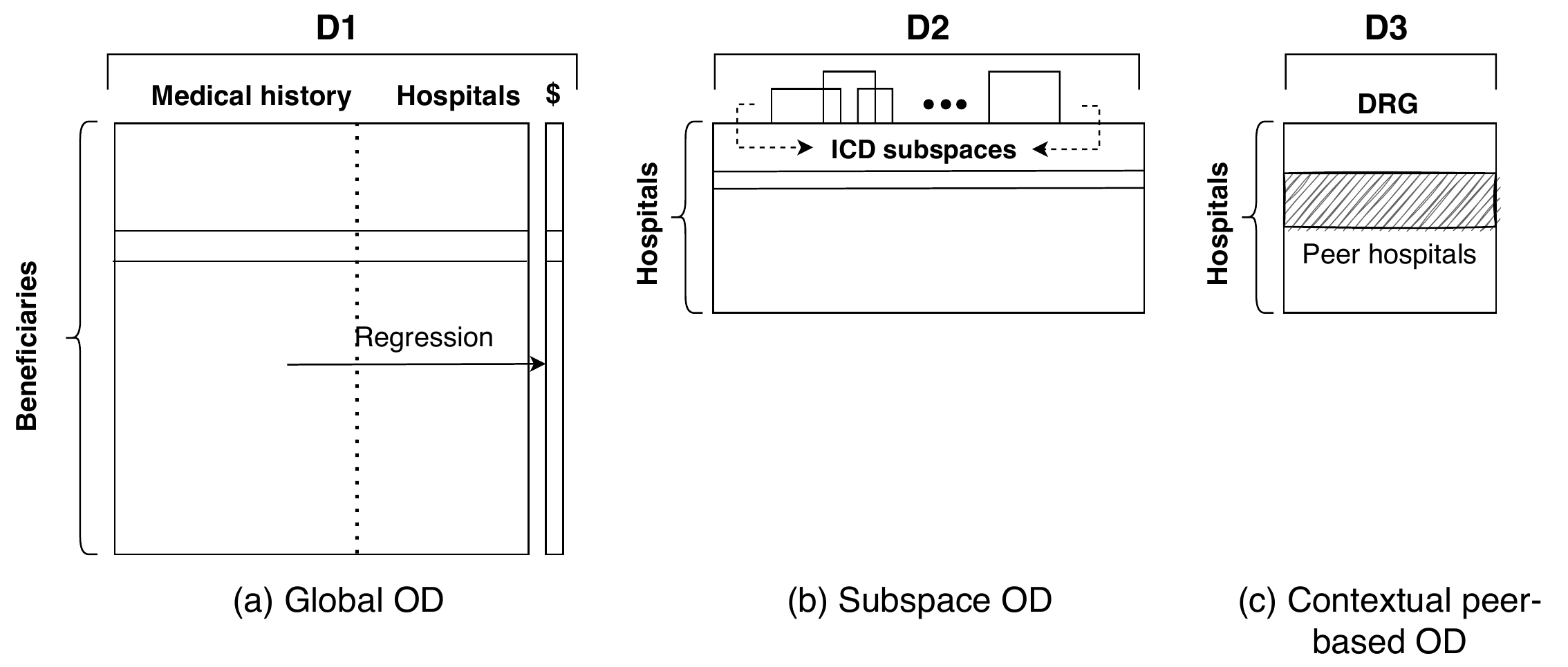}
    \caption{Multi-view anomaly detection on different \medicare data modalities -- D1, D2, and D3. Model (a): Global detector based on fixed effects regression model. The coefficient of a hospital is an indicator of excess cost of care at the hospital. Model (b): Local (in ICD codes) detector in the very high dimensional ICD code frequency representation of hospitals. It explains anomalies based on feature importance, i.e. with respect to specific ICD codes. (c): Local and contextual (peer-based) detector based on comparing DRG frequency distributions. It provides a contrastive explanation in terms of excess cost of treatment when compared to peers.}
    \label{fig:multiviewdata}
\end{figure}

Ensemble methods utilize multiple base detectors, where under certain accuracy and diversity conditions, they are
to obtain better performance than the constituent base detector alone and produce more robust results \citep{aggarwal2017outlier}. Diversity is an important property of ensemble methods, which ensures that the base detectors make independent errors that cancel out when aggregated. Therefore, various approaches have been proposed toward promoting ensemble diversity \citep{kuncheva2003measures,nam2021diversity}. In essence, our approach utilizes the diversity of the underlying data representations to induce diversity in the ensemble.

Figure~\ref{fig:multiviewdata} shows the different Medicare data modalities we consider and provide a high level description of the corresponding base outlier detection (OD) model that utilizes it. 
The first model (a) is set up as a global regression onto cost per beneficiary (target variable) from data (denoted \textbf{D1} on the figure) reflecting a beneficiary's medical history and the hospitals that they visited.  
The second OD model (b) performs outlier detection among hospitals as represented by the frequency of ICD codes used in their claims (denoted \textbf{D2}).  
Anomalous coding may be associated with only a few ICD codes (i.e. features) at a time, rather than all. Therefore, the second model is a feature subspace detector, finding outliers locally in subsets of features. Finally, the third OD model (c) performs contextual detection, identifying hospitals that behave differently from their peers. Behavior is captured by the frequency distribution of the DRG codes assigned to each hospital's claims (denoted \textbf{D3}). Here, we recognize the heterogeneity among hospitals and compare a hospital's behavior locally, i.e. in the context of its peers with similar characteristics. 

In addition to detection, our proposed models can provide explanations for their flagged anomalies. This is especially important in the absence of any ground-truth labels in practice, aiding sense-making, verification and decision making (such as whether to conduct additional investigation or to audit). By capitalization on different data representations, our method leads to different explanations with each OD model, enabling a multi-view reasoning. Specifically, in Figure \ref{fig:multiviewdata}, the regression coefficient associated with a hospital in our first OD model (a) would be a direct indicator of excess spending at the hospital. The second OD model (b) quantifies feature (i.e. ICD code) importance, and can explain each flagged anomalous hospital based on the specific ICD codes that they use differently in their claims.
The last OD model (c) provides contrastive explanations, through comparing DRG frequencies of a hospital to those of their peers. As the DRG code of a claim dictates cost, differences in the DRG coding distribution can be directly translated to excess cost of treatment. Importantly, the explanation can pinpoint which DRGs are most contributing to large excess cost of a hospital, facilitating auditing.


To arrive at a final anomalous ranking based on different modalities, we combine the rankings from individual detectors such that it captures the agreement among them. In effect, the ensemble approach allows us to gather evidence from multiple models, each leveraging a different data modality. Further, it can be ``unrolled'' to provide explanations to each flagged anomaly by each detector in the ensemble. Overall, such a multi-view detection and explanation approach takes advantage of corroborating evidences across modalities, and provides a multi-view perspective toward reasoning about suspicious behavior.


\hide{
The goal of the ranking is to assist in the audit process. Therefore, we require our anomaly detection algorithm to produce the final rank list with the following properties.
\begin{enumerate}
	\item Top ranked providers should correspond to the ones that engage in fraud and abuse
	\bit
	    \item Minimize audit cost on false positives
	\eit
	\item The assigned rank should be explainable 
	\bit
		\item Quantifiable in terms of \$ amount savings
	\eit
	
\end{enumerate}
}



The following three sections are organized to present the details of our detection models, in terms of data set up, detection methodology and explanation.\looseness=-1

\section{Expenditure-Based Detection with Massive Fixed-Effect Regression}
\label{sec:regression}

The goal of a provider-level analysis of expenditure is to understand which providers are associated with high spending on a beneficiary's hospitalization. The incentive of providers who commit fraud is to receive higher reimbursement, and so unexplained high expenditure is potentially suspicious. Our design detects high expenditures that are unexplained by a patient's medical history, which could reflect unnecessary or excessive billing. While any individual patient may receive entirely necessary high levels of care -- for example, in response to a severe accident -- when a provider's patient population consistently shows expensive, unexplained high expenditure, this may be indicative of fraud or waste.

Our design considers expenditure as a function of a patient's medical history. We collect each beneficiary's medical history, using claims from physicians office visits, hospital outpatient visits, and hospital inpatient visits over a five-year period before the target year. The outcome or target variable is the base claim amount per beneficiary per provider in the current year.

\subsection{Data Setup}
\subsubsection*{Base payment amount.~}
For our analysis, we use the base payment amount computed from the Medicare inpatient claims. As explained in Section~\ref{sec:background}, the Medicare Prospective Payment System adjusts the claim payment amount to include expenses due to provider variables such as patient mix, disproportionate share of low-income patients, outlier cases, and expenditure on education and research. These factors are generally external to the provider's coding choice and should be excluded from analysis. Therefore, to understand provider behavior with respect to inpatient encoding, we rely on the base payment amount. The base payment amount is calculated by subtracting the reported adjustment amount from the total claim amount. While payments are also adjusted by provider location through a geographically indexed wage, we do not control for provider wage index adjustments, because the geographical factor will be picked up when controlling for patient location in our regression. 

\begin{figure}[t]
	\centering
	\subfloat[Box plot of average \textit{total claim amount} for DRG across providers from inpatient claims]
	{\label{fig:claimamt}
		\includegraphics[width=\textwidth]{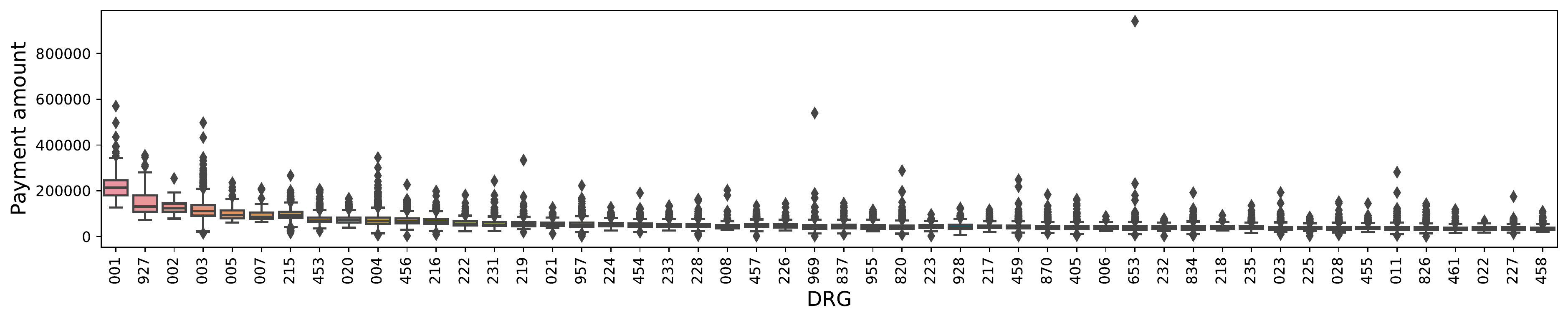}
	}\\
	\subfloat[Box plot of average \textit{base payment amount} for DRG across providers from inpatient claims]
	{\label{fig:baseamt}
		\includegraphics[width=\textwidth]{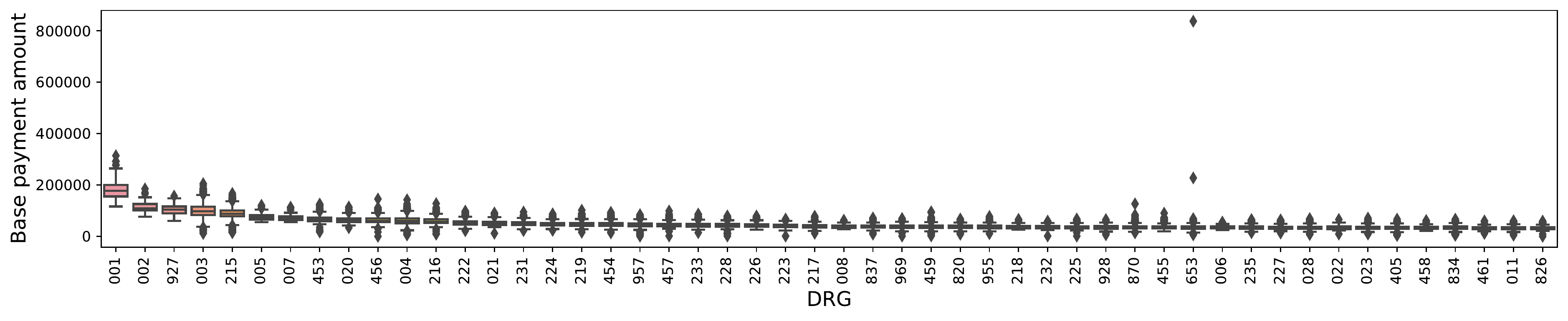}
	}
 \vspace{0.05in}
	\caption{We plot the distribution of total claim and base payment amount across providers from inpatient claims in the year $2017$. (a) Distribution of average total claim amount per provider for the top $50$ DRGs sorted on the mean of the box plot. There is large variation in the average claim amount for each DRG. (b) Box plot of the average base payment amount across providers for top 50 DRGs sorted on mean of the box plot. The variance across providers is lower for the base payment amount.}
	\label{fig:drgbox}
\end{figure}

Figure~\ref{fig:claimamt} shows the box plot of average total claim amount per provider in the inpatient claims data (year 2017) for the top $50$ DRGs, sorted by the mean of the box plot. Notice that there is large variation in the average claim amounts for each DRG. This variance across providers is reduced when the box plot instead uses the average base payment amount as shown in Figure~\ref{fig:baseamt}. However, there remains some variance across providers even when considering the base payment amount.

\subsubsection*{Patient representation.~} 
We represent each patient by their medical history and their covariates including location. 

We consider patients from $2017$ who had an inpatient hospitalization claim and are at least 70 years old.  Because Medicare is available for individuals aged 65 and older, we include patients aged $70$ years or above to ensure we observe a full 5-year history. We construct  the medical history based on a patient's provider visits in the previous five years (2012 - 2016). We filter and join patients data from physician visits, outpatient visits, and inpatient hospitalizations in the previous five years. Each patient visit, to a physician or inpatient facility, is assigned codes based on the ICD diagnosis and treatment codes. Thus, for a patient, we collect all the unique codes that were assigned in any of the visits along with their counts.

 In addition to the treatment codes, we include the chronic conditions that require regular care, associated with each patient as reported in 2016, the year before the current year. We do not include 2017 chronic conditions as those may be outcomes of the code that the hospitalizations report. Including the 2016 chronic condition of a patient helps understand any comorbidities that may arise due to their medical history and ongoing chronic condition, accounting for the increase in treatment expense. Chronic conditions include diseases such as diabetes, breast cancer, or Alzheimer's disease. Our data provide a comprehensive view of the past treatments received by a patient, and reflects on their health. Further, to account for variation due to a patient's choice of provider, as well as geographic differences in hospital reimbursement rates, we include the patient's location, represented by the first three digits of their zip code. 

\subsection{Detection Model}
To estimate expected treatment expense for a patient, we employ a fixed-effects regression model with the outcome or target variable as the total base payment, and the features being the aforementioned patient representation (medical history and location). 

We then include as regressors variables corresponding to the count of hospitalizations for that patient at each provider. The coefficients of the provider variables from this regression give the provider fixed effects -- in per hospitalization terms--  that we use to rank providers.

Note that, because we are interested in capturing the provider-level dependency of cost, we do not include treatment codes from the current year's hospitalization.  The codes of the current year's hospitalization reflect the hospital's coding decision, which can be an element in its fraud or overbilling behavior. We address those in Section \ref{sec:subspace}. Instead, the providers are added to the model to account for treatment expenses in the current year that are not reflected by the patient's medical profile; see Figure~\ref{fig:multiviewdata}(a).

\subsubsection*{Regression model specification for expenditure.~}
Given ($i$) patient representation $\bs{\x} \in \mathbb{R}^{N \times M}$ for $N$ patients, each with a $M$-dimensional representation of historical medical profile based on  the last five years (2012--2016), and ($ii$) the total base payment $\y$ in year 2017; the specification for expected treatment expenditure prediction is as follows.
\begin{align}
\label{eq:regress}
    \y_i \;=\; \beta_0 \;+\; \x_i \; \bs{\beta} \;+\; \sum_j \alpha_j\; \provider_{j,i} \;+\;  \epsilon_i \;\;,
\end{align}
where $\y_i$ is the total base payment expense for a patient $i$ in 2017; $\x_i$ is the patient representation for $i$, $\bs{\beta}$ depict regression coefficients associated with patient medical profiles and locations,  $\provider_{j,i}$ is associated with an inpatient \medicare provider $j$ which contains total count of visits to $j$ if patient $i$ visited the provider and 0 otherwise, and $\alpha_j$'s depict the provider fixed effect regression coefficients. 

\subsubsection*{Anomaly scoring.~}
In the expenditure-based regression, a coefficient
$\alpha_j$ can be interpreted as the excess treatment cost due to provider $j$ that cannot be captured by patients'  medical profile and location. As such, we can associate the magnitude and sign of this coefficient with the excess spending by a provider, and designate it as its anomaly score.

\subsection{Model Explanation}
The regression model's provider ranking in order of anomalousness is easily explainable through the coefficient values.
Specifically, each $\alpha_j$ used for scoring and ranking has the direct interpretation as the excess expenditure on treatment for a patient when visiting the provider $j$. Therefore, the fixed effects model directly quantifies the excess dollar amount impact of a particular provider, which can be used by an auditor or investigator when deciding which hospitals to investigate.

\subsection{Evaluation}
\begin{figure}[!h]
    \centering
    \includegraphics[width=0.5\linewidth]{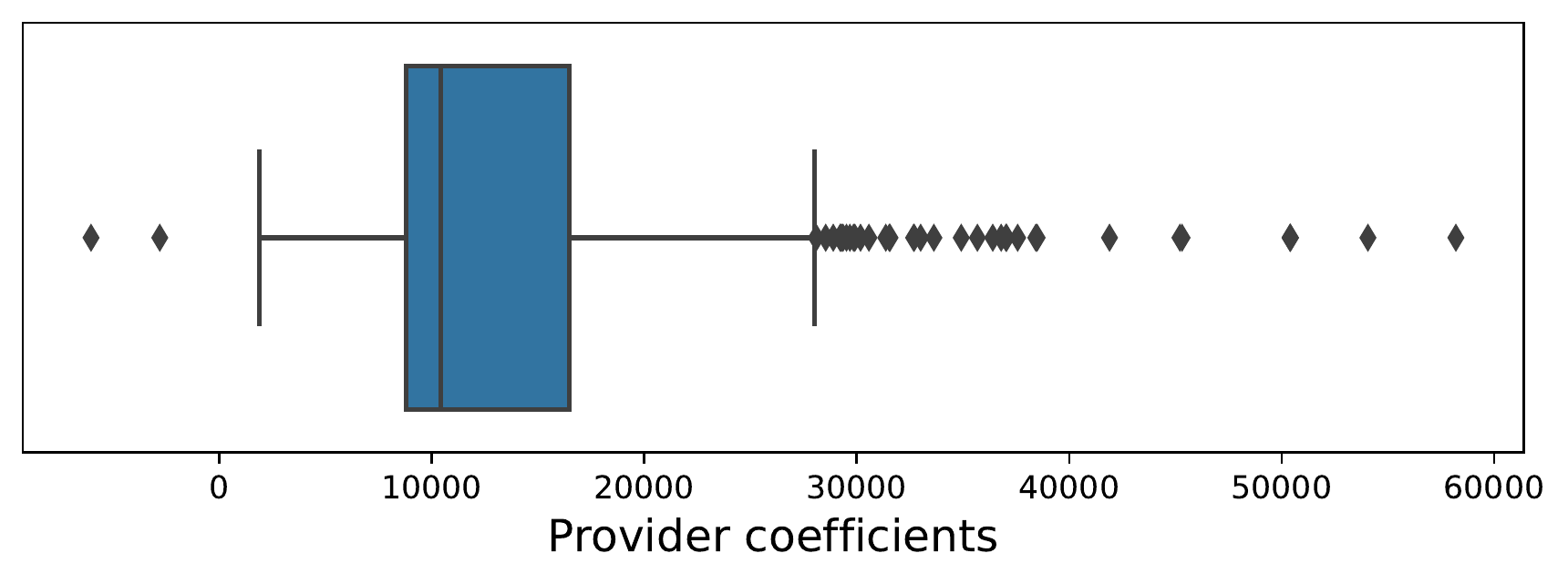}
    \vspace{-0.1in}
    \caption{Distribution of the excess cost of treatment, that is, of $\alpha_j$'s in Eq. \eqref{eq:regress} per provider $j$. The providers with large excess cost (coefficient) are ranked at the top for audit.}
    \label{fig:fixedeffects}
\end{figure}

Figure~\ref{fig:fixedeffects} shows the estimated fixed effects, i.e. the $\alpha_j$ coefficients, for providers from our expected expenditure model. The providers with large fixed effects are ranked at the top and flagged as being of  suspiciously expensive.  In auditing, it is often the case that auditors have a limited budget (time and other resources) for processing red-flags and taking action. Thus, our method allows for targeting of audits towards the most suspicious providers, which corresponds to the highest unexplained spending.

To evaluate the effectiveness of our provider ranking, we use the partial list of known fraudulent providers based on the \doj press releases described in Section \ref{ssec:antifraud}, and we compare our suspicious providers to known fraudulent providers. 
 We quantitatively evaluate the targeting of fraudulent providers using two ranking quality metrics, namely a Precision-Recall (PR) curve, and a Lift curve. 
The PR curve depicts the positive predictive value (precision) on the y-axis versus the  true positive rate (recall) on the x-axis. In audit scenarios with limited budget, a high precision at the top of the ranked list would be useful. Similarly, lift curve measures the targeting effectiveness on y-axis when compared to a random baseline as we move along varying fractions of the ranking on x-axis. 

\begin{figure}[ht]
	\centering
	\subfloat[Precision-Recall curve]
	{
	    \label{fig:apfixed}
	    \includegraphics[width=0.44\textwidth]{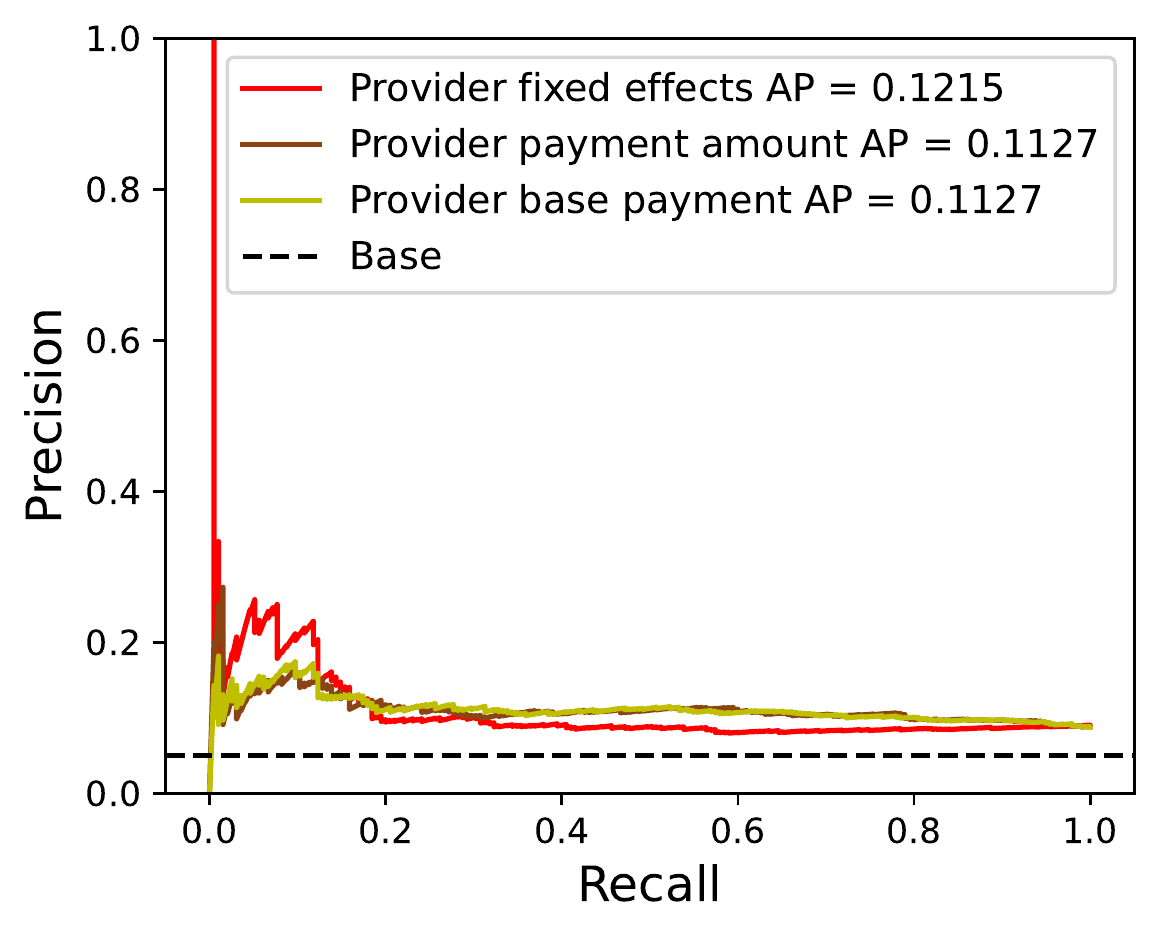}
	}
	\subfloat[Lift curve]
	{
	    \label{liftfixed}
	    \includegraphics[width=0.44\textwidth]{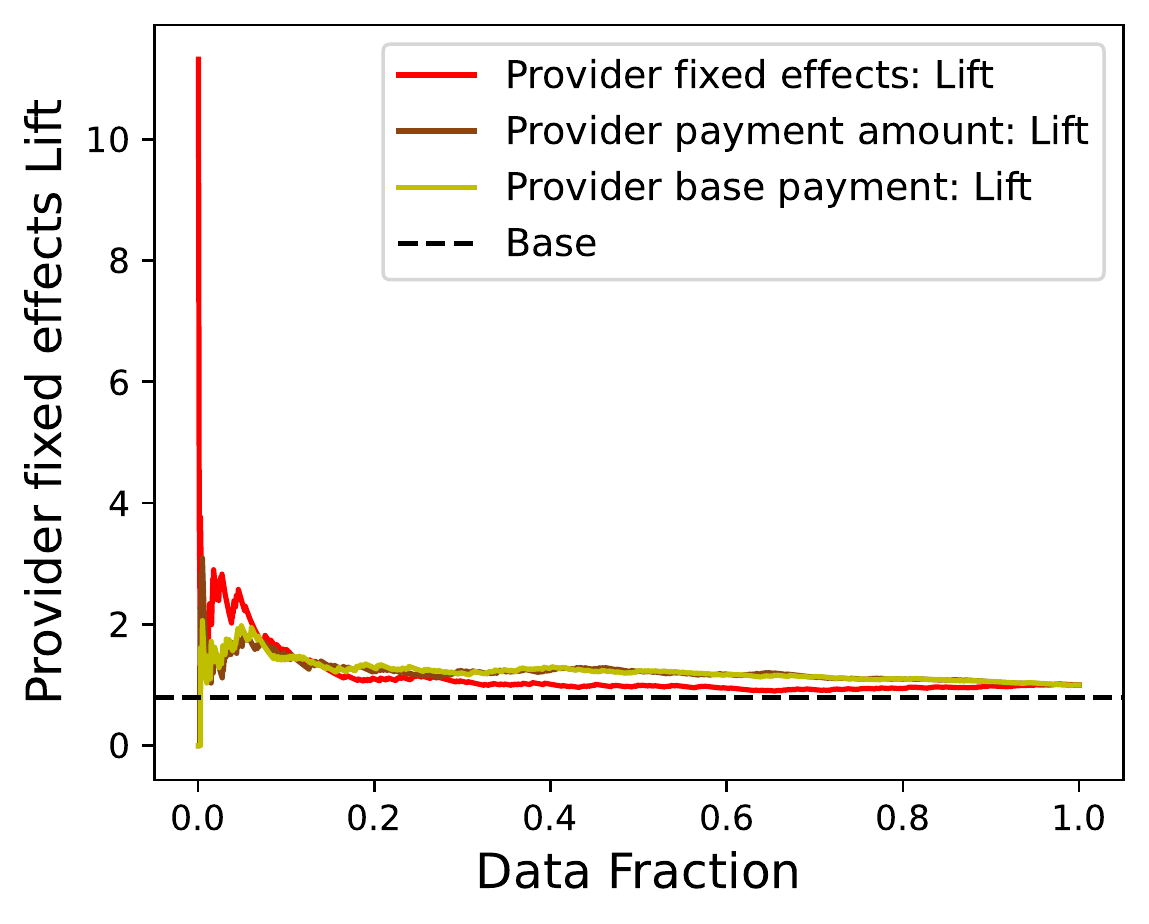}
	}
 \vspace{0.05in}
	\caption{We report (a)  Precision-Recall curve (AP: Average Precision denotes area-under-curve)  and (b)  Lift curve for provider ranking produced by fixed effects coefficients against two simple baselines: ranking of providers based on (1) average total claim amount and (2) average base payment amount. Dashed horizontal line `Base' depicts the random ranking. Notice that top of the ranking is comparatively better as evidenced by higher precision and lift when recall and top data fraction are low. This is particularly helpful for auditors who would typically process only top ranked providers  under limited budget.}
	\label{fig:fixedeffectsperf}
\end{figure}

Figure~\ref{fig:fixedeffectsperf} reports the PR and Lift curves for our fixed effects model, and compares its performance against two simple intuitive baselines. The baseline methods rank the providers based on average total claim amount and average base payment amount, respectively. Note that our fixed effects model is comparatively more effective at targeting fraudulent hospitals, with relatively higher precision and lift at the top positions.

\begin{figure}[ht]
	\centering
	    \includegraphics[width=0.5\textwidth]{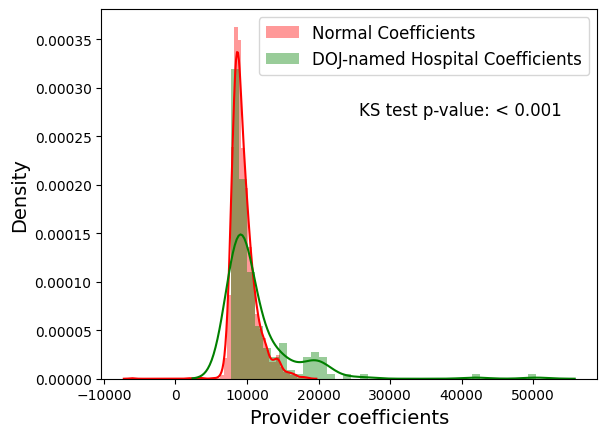}
	\caption{Comparison of fixed effect coefficients for providers facing anti-fraud lawsuits (known fraudulent entities or outliers) versus the rest of the providers (normal entities). A two-sample test rejects the null hypothesis, implying significantly different distributions statistically.}
	\label{fig:fixedeffectkstest}
\end{figure}

Figure~\ref{fig:fixedeffectkstest} reports the result of a two-sample test on the fixed effect coefficients as estimated by our model for providers in the \doj corpus versus the rest of the providers. Notice that the \doj providers typically have larger fixed effects as compared to others, and their distribution is significantly different as the test rejects the null that the two sets of coefficients are drawn from the same distribution, with $p < 0.001$.
We remark that the reported performance is conservative and only the lower limit on our model's targeting ability, since many top ranked providers that are not part of \doj ground truth may still have been involved in suspicious behavior.
We report more qualitative results, and provide case studies through explanations into such flagged providers in Section~\ref{ssec:cases}, after accounting for the evidence from other models in our ensemble.

\section{ICD Coding-Based Detection with Subspace Analysis}
\label{sec:subspace}
International Classification of Disease (ICD) codes are used by health care providers to characterize a patient's medical condition and treatment. The US uses ICD-10 codes, which were developed by the World Health Organization and can be used to designate the universe of medical issues and procedures. ICD codes encode provider assessment of a patient based on their reason of visit to the hospital and their medical conditions, and primarily reflect the diagnoses and applied procedures for treatment. For Medicare billing, the assigned \icd codes are then used as input to a ``grouper'' software used by hospital billers that assigns a diagnostic code (DRG) based on the provider findings as indicated by the assigned \icd codes. As discussed above, in the Medicare \pps, the DRG code determines the reimbursement level. Consequently, \icd coding presents opportunities for miscoding, as providers may try to achieve a more expensive DRG code to obtain higher reimbursement. Therefore, the objective of our \icd coding based analysis is to understand provider coding practices that could reveal the coding patterns applied by providers engaging in fraudulent behavior.

\subsection{Data Setup}
\subsubsection*{Provider representation.~} We use inpatient claims from the year $2017$ to understand how providers assign \icd codes to each claim, and represent providers through their reported \icd codes, including diagnostic and procedure codes. This representation captures the coding practices of a provider. 

Importantly, since providers have a choice of \icd codes, we also account for \icd \textit{code substitutability}, where a slightly similar \icd code could be used instead to yield higher reimbursements.  To capture code substitutability, 
we estimate the semantic similarity of the description of each code within each chapter of the \icd code hierarchy. Here, the description of each \icd code is constructed by concatenating its text description to the description of its ancestor codes within the \icd hierarchy. Then, pairwise Jaccard distance is computed between the descriptions of the codes and the provider representation is updated using the \icd code similarity. 

For example, the description of \icd code J45.20 under chapter X is constructed by concatenating the descriptions of J00-J99 chapter, J40-J47 block, J45, and then the \icd code J45.20 resulting in the description given as ``Diseases of the respiratory system – Chronic lower respiratory diseases – Asthma – Mild intermittent asthma uncomplicated. This representation ensures that codes with similar positions in the ICD hierarchy have somewhat similar text descriptions and are therefore near each other in Jaccard distance.


Specifically, let $\x^{\icd} \in \mathbb{R}^{\nprovider \times \nicd}$ be the matrix representation of $\nprovider$ providers in terms of $\nicd$-dimensional \icd codes in which the entries depict the total code usage count by provider,
and $\bs{J} \in \mathbb{R}^{\nicd \times \nicd}$ be the \icd substitutability matrix consisting of pairwise Jaccard similarities. Then, the provider representation $\x^{\icd_{sim}} \in \mathbb{R}^{\nprovider \times \nicd}$ after incorporating the code substitutability is given as $\x^{\icd_{sim}} = \x^{\icd} \times \bs{J}$, which re-distributes each code's frequency to substitutable \icd codes that are not directly reported in the claims data. 

We note that $\x^{\icd_{sim}}$ is very high dimensional ($> 40,000$ features). However, anomalous coding of a claim is likely covert and  associate with only a few \icd codes. Therefore, we employ a feature \textit{subspace} based detector for finding outliers locally among subsets of \icd codes. Figure~\ref{fig:multiviewdata}(b) shows this setup.

\subsection{Detection Model}
We employ a suite of subspace outlier detectors on the high dimensional provider representation $\x^{\icd_{sim}}$  to find providers deviating from the majority coding practices within certain \icd subspaces. As we are interested in \icd subspaces  that are relevant for a variety of aberrant provider practices, we utilize an ensemble of subspace detection methods that are effective on high dimensional data. In the same spirit as with our overall approach, the ensemble allows us to examine multiple diverse subspaces as each subspace detection method implements a different methodology for exploring candidate subspaces. In particular, our subspace ensemble uses five different state-of-the-art methods that we describe briefly below.\looseness=-1 



\subsubsection*{Subspace outlier detection.~}
While we represent a hospital in the high dimensional \icd space, the abnormal or aberrant behavior may be reflected only in a small, locally relevant subset of codes as pertains to stealthy behavior. Each OD algorithm in the ensemble explores local subspaces differently to provide evidence from diverse subsets. To that end, our OD model consists of the following subspace detectors:
\begin{enumerate}[label=(\roman*)]
    \item Subspace Outlier Degree (\sod)~\citep{kriegel2009outlier} locally examines each point (hospital) in the data. For each data point, it computes reference points through shared nearest neighbors. The subspace is then characterized by dimensions with low variance, lower than a provided threshold, within the identified reference set. It  records the deviation of each data point from the hyperplane spanned by the mean of the identified subspace, where outliers have larger deviation.
    \item Isolation Forest (\ifr)~\citep{liu2008isolation} builds a collection of randomized trees that approximate the density of data points in a random feature subspace characterized by paths in what are called ``isolation trees''. Each isolation tree is constructed by recursively partitioning data using a randomly chosen point in a randomly selected dimension, until the leaf of the tree contains a single data point. Shorter paths in a tree indicate sparse regions as fewer partitions lead to leaf nodes, and points belonging to each leaf at lower depth indicate outlierness in the subspace characterized by the tree path. 

     \item Robust Random Cut Forest (\rrcf)~\citep{guha2016robust}, like \ifr, also constructs an ensemble of randomized trees by recursively partitioning the data. It computes the model complexity of each tree as the sum of the bits required to store the depths of each point in the tree. An outlier is defined as a point which increases the model complexity significantly when added to the tree.
     
    \item Lightweight on-line detector (\loda)~\citep{pevny2016loda} constructs a collection of histograms on random 1-dimensional projections of the data. Each data point is then associated with the negative log-likelihood based on each histogram, and data points are ranked based on their average likelihood across the 1-D histograms.
    
    \item RS Hash (\rshash)~\citep{sathe2016subspace}, like \loda, is also an ensemble of histograms; however, it constructs a collection of grid-based histograms in randomly chosen subspaces, and grid sizes vary based on varying sample sizes of data. 
    Each data point is then scored by the number of sampled points sharing the same bin in the histogram. A sparsely populated bin is indicative of outlierness.
   
\end{enumerate}
We apply the above methods to $\x^{\icd_{sim}}$, the \icd representation of providers, and identify the providers that behave abnormally in various subspaces as explored by the algorithms.

\subsubsection*{Anomaly scoring.~} Each subspace algorithm assigns an anomaly score to each provider. 
The scores have different scale and semantics (path length, likelihood, etc.), and thus are not directly comparable across the methods. Therefore, we aggregate the ranking of providers based on individual scoring of each subspace method. We use the instant-runoff voting technique (details in Section \ref{sec:aggregate}) for rank aggregation from different subspace algorithms, and provide the final ranking of hospitals by anomalousness across all subspaces.

\subsection{Model Explanation}
We explain the ranking of a subspace detector using Shapley Additive Explanation values (SHAP values), introduced in ~\cite{lundberg2017unified} and \cite{ lundberg2020local}. SHAP values estimate feature importance by approximating the effect of removing each feature from the model as the average of differences between the predictions of a model trained with and without the respective feature. We regress the anomaly scores from a subspace detector onto the \icd representation of providers, and then estimate the SHAP values under the regression model. The feature contributions for each observation find the most important codes that affect the anomaly score significantly. This helps us find \icd codes that are contributors to a provider being ranked as an outlier.

Further, we provide dollar amount characterization of important features (\icd codes). Each \icd code is mapped to the most frequent DRG code assigned for the given \icd code within the inpatient claims. Since DRG codes are determinants of the payment for care, through this most-frequent DRG mapping, we associate dollar amount of reimbursement to \icd codes. This lends itself to understanding the dollar amount impact of an important ICD code for an anomalous provider as explained by SHAP feature importance values.\looseness=-1

\begin{figure}[t]
	\centering
	\subfloat[Precision-recall curve]
	{
	    \label{fig:apicd10}
	    \includegraphics[width=0.44\textwidth]{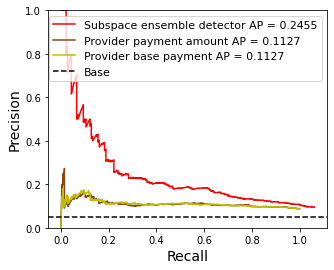}
	}
	\subfloat[Lift curve]
	{
	    \label{lifticd10}
	    \includegraphics[width=0.44\textwidth]{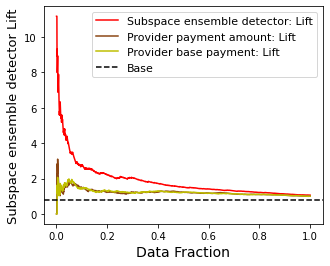}
	}
 \vspace{0.05in}
	\caption{We report  (a) Precision-Recall curve and (b) Lift curve for provider ranking produced by our ICD-10 subspace  outlier detector ensemble against two simple baselines that rank the providers  based on (1) average total claim amount and (2) average base payment amount. Dashed horizontal line `Base' depicts the random ranking.}
	\label{fig:icd10perf}
\end{figure}

\subsection{Evaluation}
Figure~\ref{fig:icd10perf} reports the performance of our subspace OD model in terms of the PR and Lift curves, using the DOJ ground truth. The subspace model ranking is at least $2\times$ better at targeting fraudulent providers compared to our two baselines, respectively based on total claim payment and base payment amounts. Our method substantially outperforms random auditing or even detection based strictly on payment amounts.

\section{Expenditure-Based Detection with Peer Analysis} 
\label{sec:peers}

Our third  model is based on peer-based excess spending detection and examines the coding decisions of hospitals as compared to similar ``peer'' hospitals that treat similar populations.  
In short, we identify hospitals who are exposed to the same patient population but manage to assign more expensive DRG billing codes.

The objective of the peer-based analysis is uncovering the local patterns of spending behavior among a {\em related} group of providers called peers, and identifying providers deviating from the group's expected behavior. We utilize the inpatient claims to create a profile for each provider under two complementing data modalities,  based on: (1) type of services provided by the hospital, and (2) the patients' chronic condition profiles served by a hospital. We then find groups of related providers based on the similarity of their provider profile representation.\looseness=-1 

To identify a locally aberrant behavior, each provider is represented in terms of its DRG frequency distribution, which determines spending. Then, the DRG representation of a given provider is compared to the summary DRG distribution of their peers. Figure~\ref{fig:multiviewdata}(c) visualizes this setup. The providers are then ranked in order of their deviation from group behavior in terms of DRG-based spending.

\subsection{Data Setup}
\subsubsection*{Provider representation.~} We construct hospital profiles to capture the nature of services provided, the characteristics of patient population served, and encoding practices that drive spending for treatment.

{\em Provider profile -- Type of services.} We first examine a provider's inpatient claims data to understand the type of services provided.  Because the DRG codes assigned by providers may be manipulated to accomplish higher reimbursement, we must not represent providers by the exact DRGs they use; instead, we consider the provider's distribution into major diagnostic categories (MDC)~\citep{mdc2022resdac}.
Each MDC corresponds typically to one major body system (circulatory, digestive, etc),
and can be associated with a set of medical specialties; each MDC contains a large set of potential DRGs. Therefore, characterizing providers by MDC allows us to consider providers that treat patients with similar types of medical needs, but without relying on the exact DRG codes assigned. For each provider, we record the normalized count of each MDC code in the inpatient claims data in the current year.

{\em Provider profile -- Patient population.} We create another profile based on patient population characteristics served by a provider. The underlying motivation for this profile is that two providers should be similar if they serve patients with similar medical conditions. To characterize the patient population at a broad level, we use the underlying chronic conditions of the patients. The chronic conditions flag whether a patient has received a previous set of services related to a chronic condition such as diabetes or ischemic heart disease. As a provider's representation, we record the normalized count of the chronic conditions of all the patients treated at the provider.

{\em Provider profile -- Spending for care.} 
The spending amount in each claim is directly tied to the assigned DRG code. To capture the DRG encoding practices of a provider, we represent each provider using the normalized counts of DRG codes from its inpatient claims. The DRG frequency representation  allows us to compare and contrast the spending between a hospital and its peers that provide similar services or serve similar patients.


\subsection{Detection Model}
\subsubsection*{Peer identification.~}
We create peer groups of hospitals that share similarities in the type of services provided or the patient population served.

Let $\pvec_j$ denote the representation for provider $j$; either based on the type of services profile using MDC codes or based on the patient population profile using chronic conditions of patients. 
We note that the provider representations are frequency distributions, as they depict normalized counts. Therefore, 
to measure the similarity between two providers $j$ and $k$, we use the Hellinger distance for probability distributions, which is an upper bound on the total variation distance~\citep{BarYossefJKS2004}, given as
\begin{align}
    d_{jk} = \frac{1}{\sqrt{2}} \cdot \lVert \sqrt{\pvec_j} - \sqrt{\pvec_k}\rVert_2
\end{align}
We examine the distribution of pairwise similarity values to decide on a  threshold $\tau$ to include only the most similar providers in a provider's peer group. 

For each provider $j$, the providers with similarity to $j$ above  $\tau$  constitute $j$'s peers, denoted $\mathcal{P}_j$. Notice that the peers are specified for each provider separately, rather than using any clustering algorithm.
This allows us to create compact peer groups of varying sizes. 
We note that fixing the peer group size would be a subpar alternative, since $j$'s group may then include distant providers as peers, skewing the representative summary statistics of the group that $j$ is compared to.



\subsubsection*{Anomaly scoring.~}

In the Medicare \pps, the reimbursement amount for treatment is directly based on the assigned DRG code to a claim. Therefore, for anomaly scoring, we utilize the provider representations over DRG codes from the inpatient claims, which consist of the normalized counts of the DRG codes used by a provider. In short, this detection mechanism assumes that providers who treat similar patient populations, or provide care for similar illnesses and injuries, should have similar DRG distributions.

For each provider, we have identified a group of providers (peers) with similar characteristics---type of services provided and patient population served---based on which we create a peer group summary in terms of distribution over DRG codes. The summary distribution is created by incorporating DRG frequencies from all the peers, weighted by their similarity to the provider under consideration. Let $\pvec^{DRG}_j$ be the DRG distribution for provider $j$ with $n_j$ claims, and $\qvec^{DRG}_j$ be the summary DRG distribution based on provider $j$'s peers, 
defined as follows.
\begin{align}
\label{eq:peerexcess}
\resizebox{0.94\hsize}{!}{%
$
\qvec^{DRG}_j = \frac{1}{Z}\sum_{k \in \mathcal{P}_j} n_k \times (1 - d_{jk}) \times \pvec^{DRG}_k \;\;\; \text{where} \;\;  \mathcal{P}_j = \{ \; k \;\vert\; (1 - d_{jk}) \geq \tau \;\} \; \text{ and }
Z = \sum_{k \in \mathcal{P}_j} n_k \times (1 - d_{jk})
$}
\end{align}
Next we tie the DRG usage frequencies to average dollar amount spending by Medicare, as the former dictates the latter. $\cost(c)$ denotes the average base price of DRG code $c$ computed from the inpatient claims data from the year $2017$. Then, the excess spending for treatment per claim on average for provider $j$ is given as follows:
\begin{align}
    \excess_j \;=\; \sum_{c \in DRGs} \; \cost(c) \;\times\; (\pvec^{DRG}_{j, \;\text{index}(c)} - \qvec^{DRG}_{j,\; \text{index}(c)})
\end{align}
where $\pvec^{DRG}_{j, \;\text{index}(c)}$ is the frequency corresponding to DRG code $c$ in the DRG representation $\pvec^{DRG}_j$ for provider $j$, and $\qvec^{DRG}_{j,\; \text{index}(c)}$ denotes that for DRG code $c$ in the peer group summary representation $\qvec^{DRG}_{j}$. In short, this amount computes how much more a provider spends because they use a different set of DRG codes than their peers, based on the average price of those DRGs.

The calculated $\excess$ amount
is the anomaly score based on which the providers are ranked, 
as it  depicts the average spending discrepancy for a provider when compared to  peers of the given provider. Since we create two different peer groupings -- one based on services provided, and another based on patients served -- we obtain two rankings, later combined through instant-runoff voting (Section \ref{sec:aggregate}).


\begin{figure}
    \centering
    \includegraphics[scale=0.54]{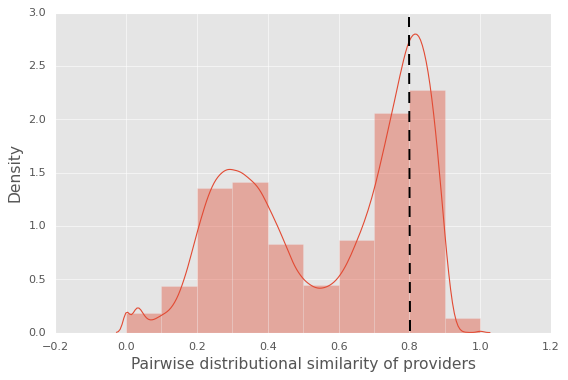}
    \caption{Distribution of pairwise similarities between provider representations. A provider and its peer hospital pair has similarity $>= 0.8$.}
    \label{fig:simdist}
\end{figure}

\subsection{Model Explanation}
The peer based OD model's anomaly score is the estimated excess spending, which is directly interpretable as the extra dollar amount a provider charges on each claim on average as compared to what would be expected from other similar providers.  
Further  explanation can be provided for a top-ranked provider by contrasting their frequency distribution over DRG codes against their peers. This allows auditors to have a contrastive understanding of DRG codes used by similar providers, and to pinpoint to specific DRGs with large frequency discrepancies.  
Direct usage comparison of individual DRGs could  point to specific codes that contribute most to the overall spending at a provider, and guide a deeper investigation of the claims associated with those specific DRG codes.

\subsection{Evaluation}
Figure~\ref{fig:simdist} shows the distribution of pairwise similarities between hospitals, and 
mark the similarity threshold at $\tau=0.8$ which is used in our implementation for identifying peers. 
We exclude providers from our analysis that have less than five peers for the chosen threshold, as the estimation of excess spending could be noisy for these providers due to small peer group. 
Providers with large excess spending are ranked at the top and are identified as suspicious.

\begin{figure}[h]
	\centering
	\subfloat[Precision-recall curve]
	{
	    \label{fig:appeer}
	    \includegraphics[width=0.44\textwidth]{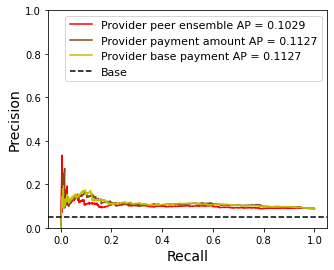}
	}
	\subfloat[Lift curve]
	{
	    \label{liftpeer}
	    \includegraphics[width=0.44\textwidth]{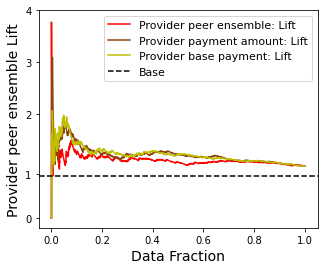}
	}
 \vspace{0.05in}
	\caption{We report the performance of ranking based on excess spending amount compared to the peers, where peers are identified via similarity based on MDC distributions and patient chronic conditions.}
	\label{fig:peerperf}
\end{figure}

We use the \doj corpus to evaluate our ranking of the providers based on excess spending.
Figure~\ref{fig:peerperf} reports the PR and Lift curves for our peer analysis. The ranking is also compared to the two baselines, respectively ranking providers by average total claim amount and average base payment amount. Although the peer-based ranking performance is comparable to these simple baselines, we remark that it is the lower bound on the performance. Furthermore, besides a mere ranking and unlike these simple baselines, our model can provide a nuanced explanation through DRG code frequency discrepancies, providing auditors with reasoning for potential factors driving 
the high spending. Finally, our model fundamentally identifies expensive hospitals as compared to their peers, which may be of interest to auditors interested in waste that may not rise to the level of fraud detected by the DOJ.

Through case studies in Section~\ref{sec:aggregate}, we report further qualitative results and provide peer-based explanations and insights into
top flagged providers after aggregating evidences from different OD models.

\section{Aggregate Provider Ranking}
\label{sec:aggregate}
Each outlier detection model presented above is a component of our ensemble method that considers a different data modality and creates a ranked list of providers based on the evidence examined individually. This ensemble method is designed to handle multi-view \medicare data, where different features of the data can be used to evaluate different aspects of suspiciousness. The goal of the ensemble is a single suspiciousness ranking for all providers. 

To arrive at the final ranking for auditing, we merge multiple rank lists into a single ranking using instant-runoff voting (IRV). Our goal is to present the aggregate ranking that is most representative of the component models. IRV combines results across rankings in a way that best reflects the information contained across multiple models ~\citep{franceschini2022ranking}. 

The rank aggregation proceeds in an iterative manner, where each round utilizes the IRV procedure to find a ``winner'' (in our case, most suspicious hospital). In each round, votes are counted for each component ranking's first choice, and a hospital with a majority of votes is then ranked at top in our aggregate ranking. The rank lists across models are updated to drop the selected hospital in this round, and the IRV procedure is repeated with updated rank lists in the subsequent rounds to arrive at an aggregate ranking. 

In our implementation, we aggregate 8 different rankings across our 3 OD models; one from the regression model, five from different subspace OD algorithms, and two from the peer-based model utilizing two separate similarity measures. Next, we show the effectiveness of our final aggregate ranking for identifying fraudulent hospitals in the \medicare system through quantitative and qualitative evaluations.



\begin{figure}[h]
	\centering
	\subfloat[Precision-recall curve]
	{
	    \label{fig:apfinal}
	    \includegraphics[width=0.44\textwidth]{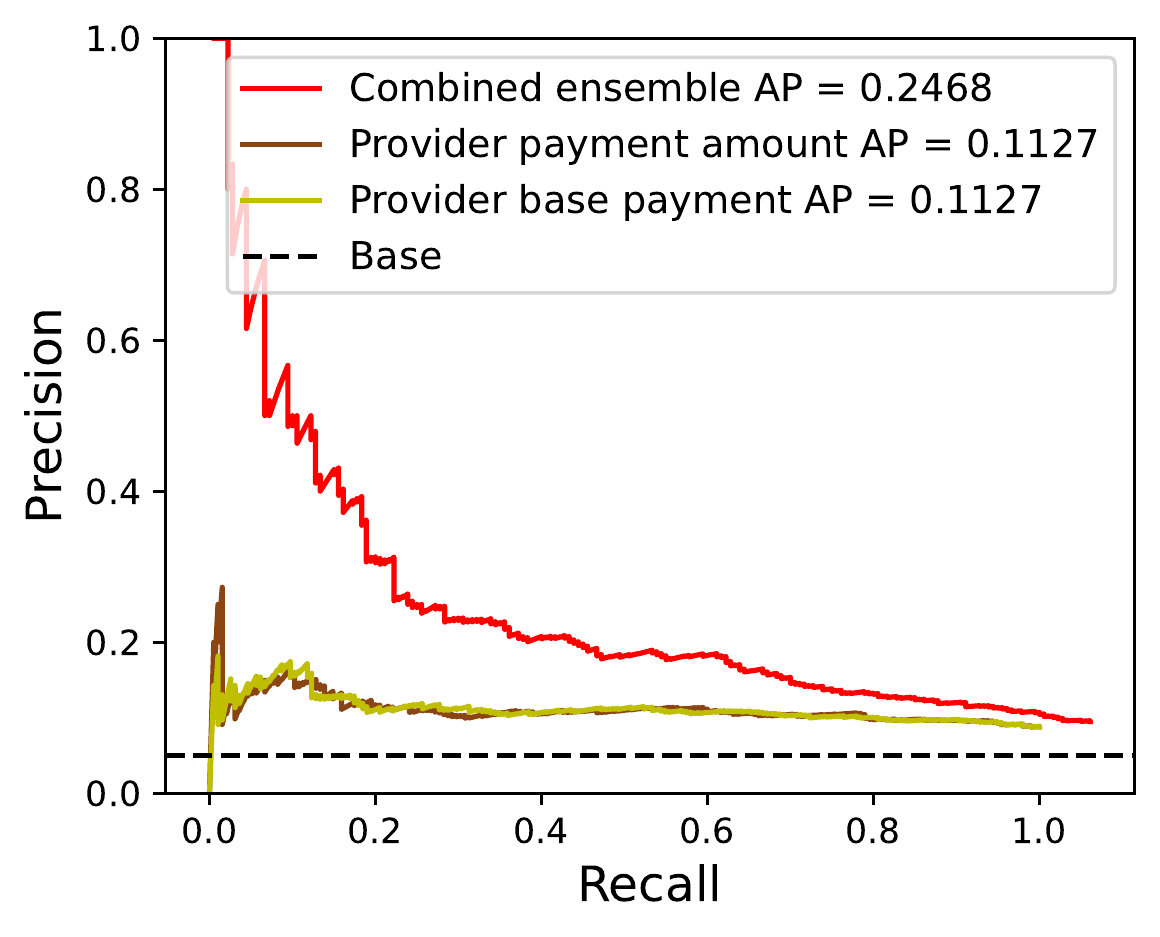}
	}
	\subfloat[Lift curve]
	{
	    \label{liftfinal}
	    \includegraphics[width=0.44\textwidth]{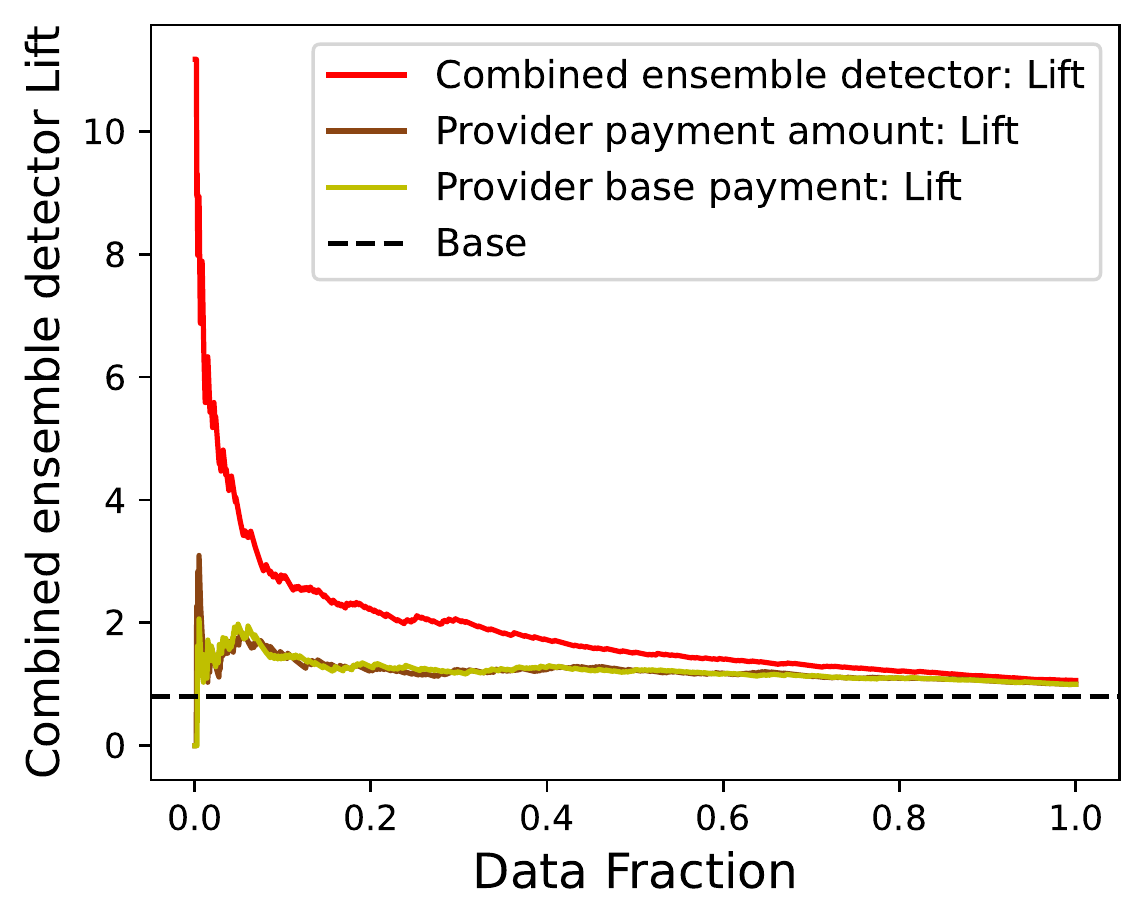}
	}
 \vspace{0.05in}
	\caption{We report the performance of the final ranking of providers as aggregated from 8 rankings based on 3 different OD models. Note that aggregated ranking improves over the ranking by individual constituent experts. The proposed ensemble is on average $4\times$ better than the random targeting of providers for auditing.}
	\label{fig:finalperf}
\end{figure}

\subsection{Quantitative Evaluation}
\label{sec:aggregateresults}
Figure~\ref{fig:finalperf} shows the evaluation of our aggregate ranking of hospitals using a PR curve and a Lift curve. The aggregate ranking is compared to intuitive baselines that rank hospitals based on their average reimbursements, or random auditing. Our aggregate ranking is able to target fraudulent providers on average {twice} as better when compared to the baseline ranking---note the area-under-curve, or average precision (AP) values on legend Figure~\ref{fig:finalperf}(a). 

While only 1 in 20 hospitals are named in the \doj Press releases, the top 50 hospitals identified by our aggregate ranking contain 21 providers named in the DOJ corpus. That is an 8-fold lift in detection rate considering the evaluation at top 50 hospitals, with an average of 4-fold lift over random/by-chance targeting across varying data fractions as seen in Figure~\ref{fig:finalperf}(b). Importantly, our ground-truth consists only of providers named in the \doj corpus, while there may be others with yet unidentified fraudulent practices -- and therefore, our list can be used to find other hospitals not yet identified as fraudulent.

\subsection{Qualitative Explanation: Case Studies}
\label{ssec:cases}
In this section, we present an analysis of our multi-view detectors, highlighting some of the salient aspects for the fraud detection task. In particular, we discuss
how our multi-view detectors can be used to explain the aberrant patterns employed by top ranked flagged hospitals
by highlighting parts of data from different views that contributed most to the ranking, which can assist in the process of auditing or deeper investigation.

We examine two top ranked providers from the aggregate ranking (1) the provider at rank 1 that is also named in the \doj corpus, and (2) the highest-ranked provider which is not in our ground truth (at rank 5, as ranks 1--4 all are part of \doj ground truth). 
In the following two case studies, we show how different models contribute evidence toward a better understanding of how each provider stands out.

\subsubsection*{Case 1: Flagged hospital named in \doj corpus}

Our aggregate ranking finds the Cleveland Clinic as the most suspicious hospital under our metrics. Here we present evidence from our 3 Outlier Detection models, where this provider is ranked at \#1 by the subspace OD model, ranked at  \#17 by the peer-based model, and ranked at \#27 by our regression-based model.

Notably, the Cleveland clinic settled with the \doj in the years 2015 and 2021 for \$1.74 million~\citep{cleveland2015settlement} and \$21 million~\citep{cleveland2021settlement}\footnote{This 2021 enforcement was against Akron General Health System, which was acquired by the Cleveland Clinic foundation in 2015.} respectively.  The evidence from our models do not directly match the reason for \doj settlements; put differently, our exact explanations have not been validated externally by litigation. Moreover, our data do not provide evidence of fraud by the Cleveland Clinic, nor do they substantiate claims from lawsuits against the Clinic. The existence of previous lawsuits by the DOJ against the Clinic validate that this is a provider with past bad behavior, and our metric indicates that this provider engaged in anomalous behavior that can be detected by our algorithm and merits deeper investigation.

\begin{figure}[!t]
	\centering
	\subfloat[Top flagged provider that is named in \doj]
	{
	    \label{fig:ccshapcc}
	    \includegraphics[width=0.5\textwidth]{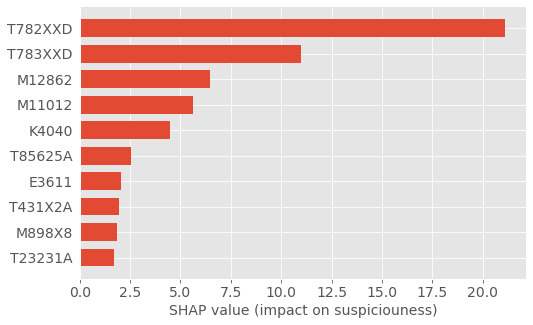}
	}
	\subfloat[Top flagged provider that is not in \doj]
	{
	    \label{fig:aoshapcc}
	    \includegraphics[width=0.5\textwidth]{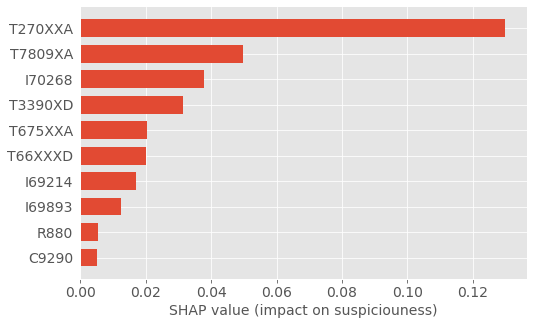}
	}
 \vspace{0.05in}
	\caption{ICD codes contributing to suspiciousness of top ranked providers based on SHAP values}
	\label{fig:shapcc}
\end{figure}

Our regression model estimates the excess expenditure on treatment for a patient when visiting the Cleveland Clinic to be \$29,844.33, which is almost $3\times$ the average expenditure ($\approx$\$10K) as shown in Figure~\ref{fig:fixedeffects}. This does not, by itself, indicate that the Cleveland Clinic engaged in bad behavior, as this may reflect that it performs more specialized medical procedures, although our regression accounts for the patient's recent medical history.

One potential concern is that the hospital highlighted in this example, the Cleveland Clinic, as particular aberrant is a unique hospital that serves a particularly sick patient pool, and that therefore, the results are driven by selection of patients into different hospitals, as opposed to the effect of being treated at that hospital on expenditure. We argue this is not the case. Indeed, the two closest peer hospitals to the Cleveland Clinic are New York Presbyterian and Beth Israel Deaconess, both of which are similarly prestigious hospitals involved in specialty care.  Therefore, we expect that the results reflect actual divergent coding patterns by the most suspicious providers, rather than detecting hospitals that are engaged in specialty treatment. 

Figure~\ref{fig:ccshapcc} plots the most important ICD codes that contribute to the anomaly score of the provider from the subspace OD model, based on SHAP values. The top ICD code ``T782XXD'' is described as ``Anaphylactic shock, unspecified, subsequent encounter'' which falls under the ancestor ``T78'' with the description: ``Adverse effects, not elsewhere classified''\footnote{ICD codes are available for lookup through ICD10Data. This code is available online at: \url{https://www.icd10data.com/ICD10CM/Codes/S00-T88/T66-T78/T78-}}. As such, T78 appears to be a catch-all classification for adverse effects for injuries, poisoning, and other consequences of external causes for visit. Moreover, the code T782XXD is considered exempt from reporting whether the condition is present on admission (POA) to an inpatient facility. The next ICD code ``T783XXD" is under the same ancestor, T78, and is also considered exempt from reporting if POA. Similarly, the description of code ``M12862'' allows non-specific reasons to be used for encoding as the given description is: ``Other specific arthropathies, not elsewhere classified, left knee''.

\begin{figure}[!t]
    \centering
    \includegraphics[scale=0.54]{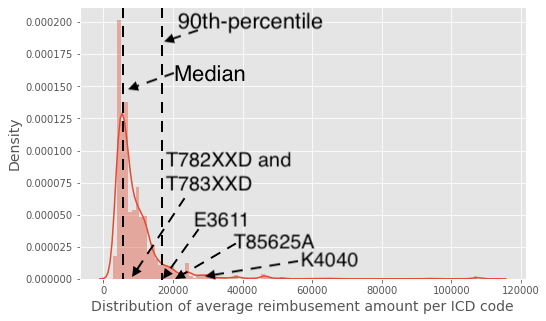}
    \caption{Distribution of ICD reimbursement amount obtained after mapping ICD code to most frequent DRG code in the inpatient claims data in year 2017. The median reimbursement amount is \$6,650.88, and the
    $90$-percentile reimbursement amount is \$16,401.04. }
    \label{fig:icdprice}
\end{figure}

We next  examine the reimbursement amounts related to these ICD codes, based on their mapping to the DRG they are most frequently associated with. The distribution of the amounts across all ICD codes is given in Figure \ref{fig:icdprice}. The codes T782XXD and T783XXD can be mapped to two DRG codes: 949 (Aftercare with cc/mcc) and 950 (Aftercare without cc/mcc).\footnote{Here, `cc' and `mcc' stand for  Complication or Comorbidity and Major Complication or Comorbidity, respectively.} The reimbursement amount for DRG code 949 is about 25\% more compared to DRG code 950, where T782XXD is reported most frequently against DRG code 949. Further, within the ICD-10 hierarchy, codes T782XXD and T783XXD are the most expensive and get at least 50\% more reimbursement than any other sibling or parent code. Notably, 6 out of top 10 ICD codes contributing to anomaly score (as shown in Figure~\ref{fig:shapcc}) have reimbursement amounts that are more than 50th percentile among all ICD codes, while 3 of them associate with DRG codes with amount above the 90th percentile (see Figure~\ref{fig:icdprice}).
All these factors explain, through specific ICD codes, associated DRGs and dollar amounts, the reasoning behind
why a flagged provider stands out. This evidence provides starting points for further investigation.

\begin{figure}[h]
\vspace{-0.05in}
	\centering
	\subfloat[Provider]
	{
	    \includegraphics[width=0.5\textwidth]{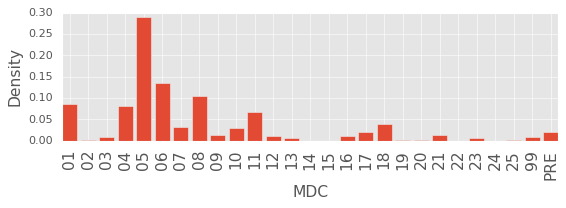}
	}
	\subfloat[Nearest peer]
	{
	    \includegraphics[width=0.5\textwidth]{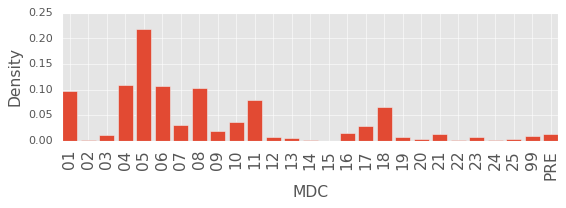}
	}
 \vspace{0.05in}
	\caption{Provider (named in \doj) and its nearest peer represented in terms of MDC codes indicating provider facilities and services provided.}
	\label{fig:mdcpeervec}
\end{figure}

\begin{figure}[ht]
	\centering
	\subfloat[Provider (named in \doj) DRG representation for MDC 05]
	{
	    \includegraphics[width=\textwidth]{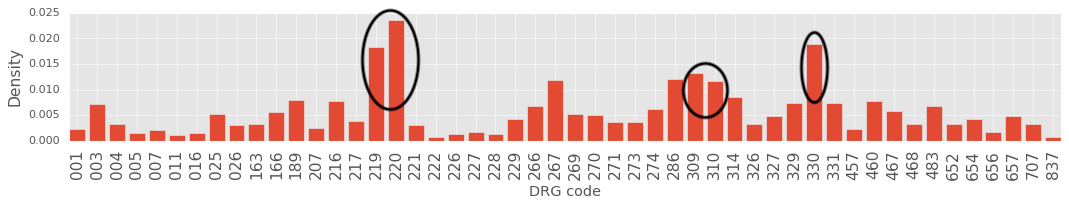}
	}
	
	\subfloat[Summary DRG distribution of its peers for MDC 05]
	{
	    \includegraphics[width=\textwidth]{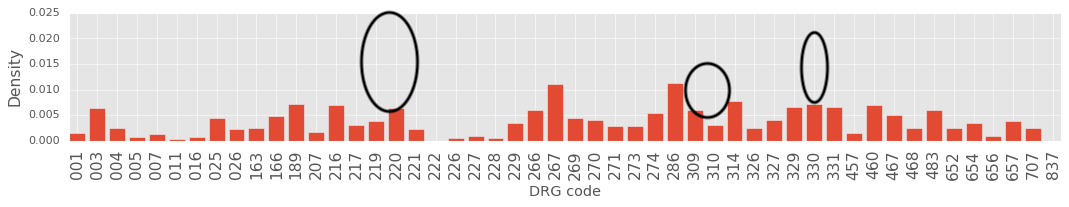}
	}
 \vspace{0.05in}
	\caption{Comparing the DRG distribution of provider (named in \doj) to the summary distribution created from its peer hospitals.}
	\label{fig:mdc05drg}
\end{figure}

In the peer-based model, the provider is flagged through the peer relation of providers with respect to their MDC representation. Figure~\ref{fig:mdcpeervec} shows the MDC distribution of the Cleveland Clinic and its nearest peer provider. Notice that in terms of facilities and services provided as encoded by their MDC, the two hospitals are quite similar. 
We compare the DRG representation of the Cleveland Clinic to the summary DRG representation of all its peer hospitals over the top 50 DRG codes that are selected based on their contribution to excess spending (see Eq.~\ref{eq:peerexcess} for excess spending estimate).
As shown in Figure~\ref{fig:mdc05drg}, Cleveland Clinic uses certain DRG codes  more frequently than its peers as indicated by the summary distribution---starting with 219, 220, as well as 309, 310, 330. DRG codes 219 and 220 belong to ``Cardiac Valve and Other Major Cardiothoracic Procedures'' with reimbursement amount in top 4 most expensive within MDC 05. DRG codes 309, 310 are described as ``Cardiac Arrhythmia and Conduction Disorders'', and DRG code 330 is described as ``Major small and large bowel procedures with cc''. Note that the description of codes 309, 310 and 330 is specific to a particular condition, while the description for 219--220 allows for ambiguity. Ambiguity may provide opportunities for miscoding to reach for higher reimbursement.

In summary, all three outlier detection models point to evidence from different views of the claims data that makes the top ranked hospital stand out from others, both in terms of local and global analysis. These pieces of evidence explain the ranking by shedding light into certain coding practices that a provider engages in, and may be utilized in further audit processes.

\subsubsection*{Case 2: Flagged hospital not in \doj corpus}

We now turn to a hospital which is flagged as suspicious by our metric but was never named in a DOJ press release.

In the aggregate ranking, AdventHealth Orlando hospital is ranked at \#5 in order of suspiciousness. All 4 hospitals higher in the ranking were named in the DOJ corpus, motivating this case study. This provider is ranked at \#5 by the subspace OD model, and ranked at \#35 by the peers-based model. 

It is important to note that our model does not provide evidence of fraud, nor do we claim that AdventHealth Orlando has committed any fraud. Rather, our ranking of hospital suspiciousness can be used to guide further investigation and audits, and we use this case study to examine how our explainable model can help direct investigatory resources toward the exact claims that make a provider different from its peers. 

Figure~\ref{fig:aoshapcc} presents the bar plot of the top 10 ICD codes by importance for the provider, based on SHAP values for the anomaly ranking from our subspace OD model. Note that 5 out of these top 10 ICD codes fall under ICD-10 chapter ``S00-T88 Injury, poisoning and certain other consequences of external causes''. The ICD code T270XXA is most frequently mapped to DRG code 205 which is described as ``Other respiratory system diagnoses with mcc''. The 3rd ranked ICD code ``I70268'' is described as 
``Atherosclerosis of native arteries of extremities with gangrene, other extremity''. Based on the descriptions of these top ICD codes, a common thread appears to be that the codes leave room for ambiguity---due to the catch-all word `other' in their descriptions. 
Further, 7 out of 10 ICD codes have reimbursement amount larger than the 50th percentile, and 4 out of 10 have reimbursements larger than 90th-percentile reimbursements across all ICD codes (recall Figure~\ref{fig:icdprice} for the ICD price distribution). 

\begin{figure}[!t]
	\centering
	\subfloat[Provider]
	{
	    \includegraphics[width=0.5\textwidth]{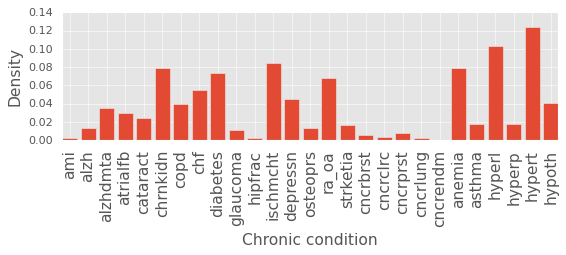}
	}
	\subfloat[Nearest peer]
	{
	    \includegraphics[width=0.5\textwidth]{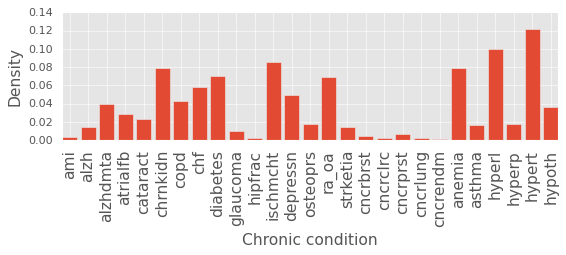}
	}
 \vspace{0.05in}
	\caption{Provider (not in \doj corpus) and its nearest peer represented in terms of patient population served}
	\label{fig:chrpeervec}
\end{figure}
\begin{figure}[!th]
	\centering
	\subfloat[Provider (not named in \doj) DRG representation.]
	{
	    \includegraphics[width=\textwidth]{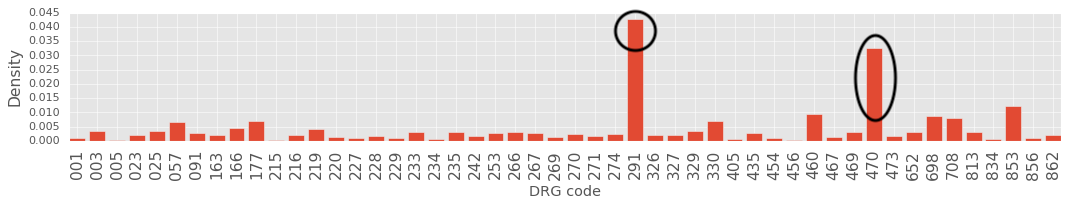}
	}
	
	\subfloat[Summary DRG distribution of its peers]
	{
	    \includegraphics[width=\textwidth]{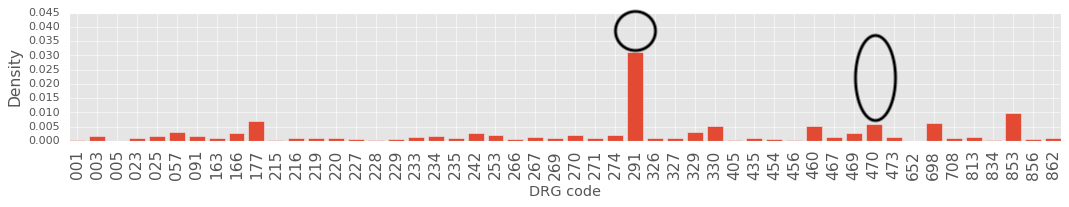}
	}
 \vspace{0.05in}
	\caption{Comparing the DRG distribution of provider (not named in \doj) to the summary distribution created from its peer hospitals.}
	\label{fig:chrdrg}
\end{figure}

Next we present  evidence from the peer-based OD model, though the provider is not top ranked in this model. Figure~\ref{fig:chrpeervec} shows the provider and its nearest peer hospital that serve similar patient populations, represented in terms of chronic conditions of the patients. We note the almost identical distributions of chronic conditions for the provider and its nearest peer hospital. We compare the DRG distribution of the provider to the summary DRG distribution of its peers. 

Figure~\ref{fig:chrdrg} shows the distribution over the top 50 DRG codes, 
where the provider's distribution deviated from the summary distribution the most weighted by DRG reimbursement amount (see Eq.~\ref{eq:peerexcess}).
We find that excess expenditure is almost entirely driven by two DRG codes, namely 291 (heart failure and shock with mcc) and 470 (major joint replacement or reattachment of lower extremity without mcc) with reimbursement costs larger than the 50th-percentile among DRG codes.

Similar to the earlier case, our models pinpoint specific ICD and DRG codes that can help jump-start further investigation, while highlighting dollar amount discrepancies that provide perspective with respect to monetary value.


\section{Characterizing Outlier Providers}
\label{sec:desc}

In this section, we examine the covariates of hospitals to understand the factors that characterize an outlier hospital as detected by our model. The covariates used in the analysis depict various hospital characteristics such as hospital rating, number of unique patients served, ownership type, location, and length of stay for an inpatient visit. These features are derived from publicly available information for all \medicare hospitals and importantly are \textit{not} included in the data used for detection. 

Understanding the factors that drive outlier provider behavior is crucial for improving the health care sector. Extensive policy reforms seek to shape the structure of the health care market, increasing regulations on providers deemed to be harmful or inefficient. By characterizing the nature of hospitals deemed suspicious by our metrics, we hope to contribute to the ongoing literature that evaluates how various interventions -- for example, those targeting for-profit care -- can affect fraudulent behavior.

\begin{figure}[ht]
	\centering
	\subfloat[Hospital rating (scale 1 to 5)\label{subfig:chr1a}]
	{
	    \includegraphics[height=1.1in]{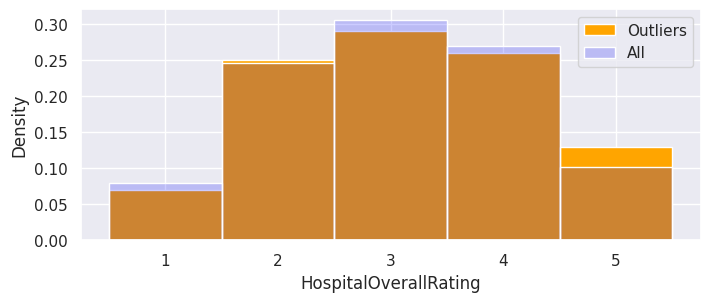}
	}
	\subfloat[Hospital ownership type: $\;\;$ Private, Govt, Non-profit\label{subfig:chr1b}]
	{
	    \includegraphics[height=1.1in]{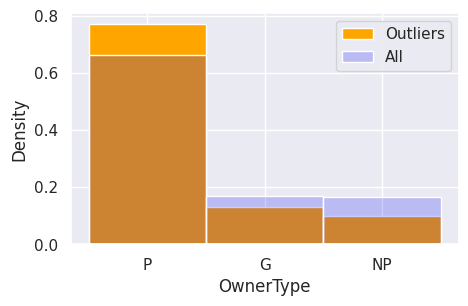}
	}
        \subfloat[Hospital location  \label{subfig:chr1c}]
	{
	    \includegraphics[height=1.1in]{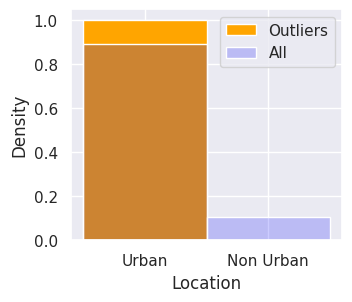}
	}
 \vspace{0.05in}
	\caption{Comparison of distributions over categorical covariates for Outlier hospitals and All hospitals}
	\label{fig:chr1}
\end{figure}

\begin{figure}[ht]
	\centering
	\includegraphics[width=0.95\textwidth]{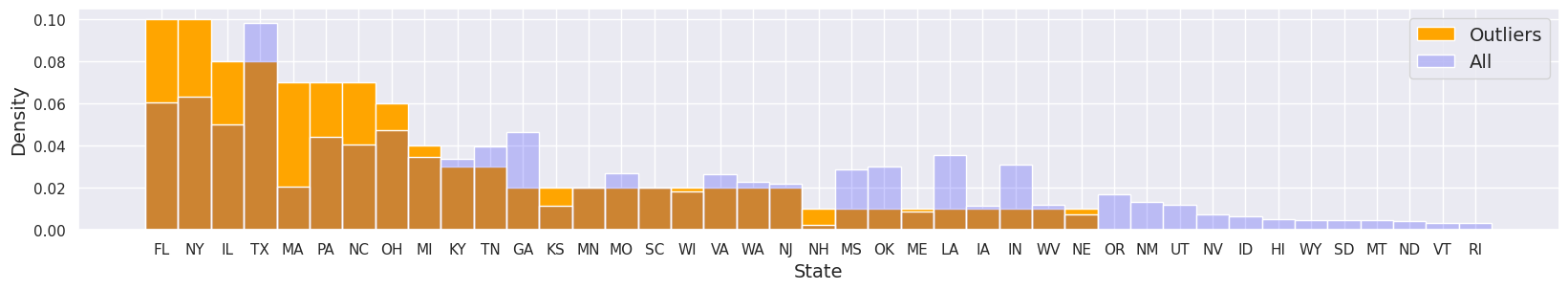}
 \vspace{0.05in}
	\caption{Comparison of distributions over `State' for Outlier hospitals and All hospitals}
	\label{fig:chr2}
\end{figure}

Figure~\ref{fig:chr1} shows the normalized histograms for categorical covariates -- hospital rating, ownership type, location type -- for the providers. We compare the distributions for the top 5\%\ of suspicious providers in aggregate outlier ranking with those over all the providers.  The idea is that, assuming fraud is rare, an investigator with limited resources would examine only the top portion of the ranked providers. 

We observe in Figure~\ref{subfig:chr1a} that histograms for Hospital Overall Rating largely overlap, indicating that outlier hospitals and all the hospitals are sampled from a similar underlying distribution, i.e. hospital rating is \textit{not} a strong predictor of outlier status.
On the other hand,  Figures~\ref{subfig:chr1b} and  ~\ref{subfig:chr1c} show that our top ranked fraudulent providers are more likely to be private (for-profit) urban hospitals, and less likely to be non-urban, government-owned or nonprofit hospitals. This observation agrees with the
literature on for-profit care, which has found distortions from this ownership structure~\citep{gupta2021does}. 

Figure~\ref{fig:chr2} compares the distributions over states where a hospital is located.  Outlier providers are more likely to be from states Florida, New York, Illinois, and Massachusetts, and less likely to be from Texas and Georgia. This is also corroborated by the \doj cases, where about $15\%$ of the named hospitals are based in Florida.
 
\begin{figure}[ht]
	\centering
        \subfloat[average Length of Stay (aLOS)]
	{
	    \includegraphics[width=0.44\textwidth]{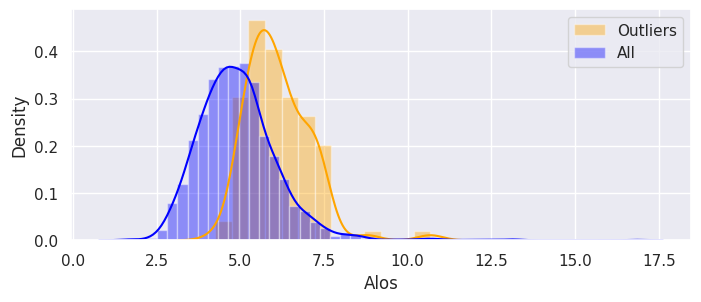}
	}
        \subfloat[\#Unique patients served]
	{
	    \includegraphics[width=0.45\textwidth]{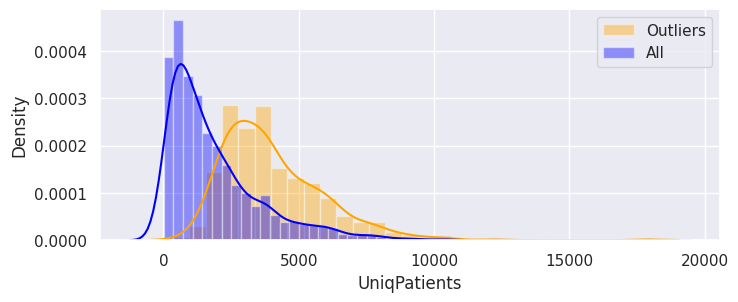}
	}
 \vspace{0.05in}
	\caption{Comparison of distributions over numeric covariates for Outlier hospitals and All hospitals}
	\label{fig:chr3}
\end{figure}

Figure~\ref{fig:chr3} compares the distributions across average length of stay and  number of unique patients served. Ranked outlier hospitals keep inpatients longer as compared to other hospitals. This could be to justify the usage of costlier DRGs, or driven by ranked outlier hospitals receiving sicker patients; however, our metrics control for patient health characteristics. Additionally, top ranked fraudulent providers serve more unique patients on average. Since a large fraction of our top ranked providers are also named by the \doj, it may indicate that a greater number of unique patients may provide more opportunity for perturbations in diagnosis coding resulting in higher reimbursements, or it could reflect the fact that our outliers are largely urban hospitals.





\section{Conclusion}
\label{sec:concl}

The unsupervised ensemble method introduced in this work provides a new data-driven approach to identifying health care fraud using massive claims data. Our approach uses different data modalities -- including patient medical history, provider coding patterns, and provider spending -- to detect anomalous behavior consistent with fraud and abuse. Besides detection, the methodology offers interpretability,  where qualitative case studies of our results based on model-specific explanations pinpoint specific ICD and DRG codes associated with excess spending at a provider. Finally, our method allows us to characterize the types of providers most likely to be ranked as suspicious, which may be useful for guiding anti-fraud policy more broadly.

Our method substantially outperforms baseline algorithms. We combine evidence from multiple unsupervised outlier detection algorithms that use different types of global and local analysis to create a final ranking of suspiciousness. While only 1 in 20 hospitals are named in our ground truth data as fraudulent, 21 of our top ranked 50 hospitals are in the same corpus, achieving an 8-fold improvement in detection rate.

Our data come from Medicare, the largest federal health care program, and we validate our method quantitatively using Department of Justice (\doj) press releases that name hospitals. Medicare spends over a hundred billion dollars per year on hospitalizations, and the federal government has limited enforcement capacity. We believe our findings are {\it per se} interesting, because they help pinpoint fraud by private firms against the government in a way that could be used to improve public spending. 

Our method has natural extensions beyond Medicare and beyond hospitalizations. We believe that the same method will prove useful in detecting fraud against private insurers, who face many of the same issues. Private insurers spend hundreds of billions of dollars per year on reimbursing care, and even small shares of fraud can be very expensive. Our detection algorithm can be used to guide auditing by identifying which providers are committing the most egregious behavior. Because our method explains which patterns drive the detection, it can facilitate auditing once a provider is selected by allowing an investigator to focus on certain billing codes and types of care. Our method also has a natural extension to Medicaid, the federal-state partnered low-income subsidy program, which spends an additional \$400 Billion per year on health care. 
 With health care spending at 19.7\% of US GDP~\cite{center2022nhe}, tools for detecting health care fraud can find wide-ranging use.

\newpage
\bibliographystyle{informs2014} 
\bibliography{BIB/ref} 


\newpage
\appendix
\section*{Appendices: For Online Publication}
\vspace{1cm}
\section{Data Preprocessing}
\label{appendix:medicaredata}
Our analysis of provider behavior uses data from each hospitalization and patient in the Medicare system. We consider patients hospitalized in 2017, and we use data from 2012 through 2016 to construct the patients' medical history.

\subsection{Processing inpatient hospitalizations}
We use 100\% of samples of Fee-For-Service inpatient claims file from the Medicare data. Annual files contain beneficiary hospitalization details including provider, assigned \drg , assigned \icd codes, and payment reimbursement details including total payment amount, disproportionate payment, education payment, and outlier amount.
The raw data is filtered to include claims where the total payment is greater than individual components. For example, if a claim has higher disproportionate payment compared to total payment amount, we exclude such a claim record from our data. These claims may indicate corrupted or noisy data recording.
Next, to meet cell-size suppression requirement under our data agreement, we exclude providers along with their claim records, who served 10 or fewer beneficiaries in 2017. 
We then create lists of unique providers and beneficiaries from the filtered data, which we utilize for merging with other Medicare files.

\subsection{Provider profile}
First, we merge the filtered data with the master beneficiary summary files which contain beneficiary enrollment information including the beneficiary's address, demographics, and chronic conditions. Next, the data are merged with a \drg to \mdc mapping. 

We then create three types of provider representations. First, we collect the counts for each unique \icd code used by a given provider, creating a representation in terms of \icd codes used. This is a very high dimensional representation, where we apply our subspace based methods.

Next, for each provider, the counts of unique \mdc codes are recorded. Since, each \mdc typically corresponds to a part of the body, the \mdc representation of providers gives a summary distribution in terms of the type of care they provide. Further, we collect counts of chronic conditions for each provider, which represents the distribution of patient population being served by a provider.

We also create the distribution over \drg codes for each provider by collecting the counts of unique \drg codes used by providers. This representation allows us to understand the spending pattern of a provider, since under the \pps system, the \drg code is directly tied to spending amount in each claim.

\subsection{Beneficiary medical profile}
In order to create a beneficiary's medical profile, we stitch through the patient's health care claims across different touchpoints in the Medicare system over the 5 years preceding the 2017 hospitalization (2012 -- 2016). 
Specifically, for these years, we use  100\% of samples of Fee-For-Service inpatient and outpatient claims, and 20\% of  samples of carrier files, which describe physician office visits. 20\% is the largest available size of carrier files.

Given the volume of the datasets, we first filter the patient's visits across datasets based on the unique beneficiary list created from inpatient hospitalizations in year 2017. For each type of visit i.e. physician, outpatient, inpatient, we find unique diagnosis codes across five years. Next, for a given beneficiary, we collect the counts over the last five years for each of the unique diagnosis codes.
We also include chronic conditions from the year 2016 and the patient's zip code from the master beneficiary summary file. Thus, a beneficiary is represented in terms of assigned codes from past visits, chronic conditions and zip code.

\section{\doj Corpus}
\label{appendix:dojdata}
We scrape and download press releases containing the word `Medicare' from the central \doj and the Offices of the United States Attorneys (USAO), which reflect local \doj branches. The base URLs used in scraping for the \doj and USAO are \url{https://www.justice.gov/news?keys=medicare} and \url{https://www.justice.gov/usao/pressreleases?keys=medicare}\footnote{ Webpages were accessed on accessed Mar 21, 2022.} respectively.

Next, we obtain the list of inpatient hospitals in the Medicare system from `Medicare Inpatient Hospitals – by Geography and Service' dataset available from the Centers for Medicare and Medicaid Services at \url{https://data.cms.gov/provider-summary-by-type-of-service/medicare-inpatient-hospitals/medicare-inpatient-hospitals-by-geography-and-service}. This contains the information on providers including name, CCN (hospital ID), city, and state. 

To find providers that are named in \doj or USAO press releases, we first run a named entity recognizer\footnote{We used off-the-shelf entity recognizer Spacy available at \url{https://spacy.io/api/entityrecognizer}} to obtain the names of all organizations from the press releases. 
We then run an exact name matching scan for each hospital in the list of Medicare inpatient providers in the recognized organizations from the press releases. Matched hospitals are then recorded as our ground truth.
Next, we also run a partial name matching. We obtain tokens for each inpatient hospital in Medicare after dropping the word ``hospital'' in their name. Then we find organizations from scraped press releases that contain the tokens for Medicare hospitals. Since, we are matching tokens, multiple organizations match for a given Medicare hospital. We manually filter the multiple match and validate the match. The ground truth is augmented with our validated matches, which forms our \doj corpus for evaluation.


\hide{
\section{Top ranked hospitals}
\label{sec:appendix}
We present the top 20 of the ranked hospitals by different components of our method. The hospitals named in \doj corpus are marked by $^*$.

\begin{table}[h]
	\caption{Ranking of hospitals by our methods in order of their suspiciousness. The hospitals named in \doj corpus are marked by $^*$. \label{tab:rankedresults}}
	\resizebox{\textwidth}{!}{
	\begin{tabular}{rp{6cm}p{6cm}p{6cm}p{6cm}}
		\toprule
		Rank & {Regression} & {Subspace} & {Peer-based} & {Aggregate Ranking} \\
		\midrule
		1. & Parkland Health And Hospital System\isout  & Cleveland Clinic\isout & St Joseph's University Medical Center & Cleveland Clinic\isout \\
		2. & Us Pain \& Spine Hospital   & New Hanover Regional Medical Center\isout & Dixie Regional Medical Center & New Hanover Regional Medical Center\isout\\
		3. & Minimally Invasive Surgery Hospital   & Tampa General Hospital\isout & South Texas Health System Behavioral\isout &Tampa General Hospital\isout\\
		4. &  Westchester Medical Center\isout  & Massachusetts General Hospital & Corpus Christi Medical Center& Massachusetts General Hospital\\
		5. & Harris Health System   & AdventHealthOrlando & Avera Sacred Heart Hospital & AdventHealth Orlando\\
		6. & Beth Israel Deaconess Medical Center\isout   & Southcoast Hospital Group, Inc & Camden Clark Medical Center & Rush University Medical Center\isout\\
		7. &  Nyu Langone Hospitals  & Lancaster General Hospital\isout & St Luke's Cornwall Hospital & Methodist Hospital\\
		8. &  New York-Presbyterian Hospital  & Beth Israel Deaconess Medical Center\isout & Faith Community Hospital & North Mississippi Medical Center\isout\\
		9. &  Mount Sinai Hospital\isout  & Orlando Health Orlando Regional Medical Center\isout & Texas Health Harris Methodist Hospital Alliance & Mount Sinai Hospital\isout\\
		10. &  Jacobi Medical Center  & Memorial Mission Hospital And Asheville Surgery Ce & Nyack Hospital & Orlando Health Orlando Regional Medical Center\isout\\
		11. &  Boston Medical Center Corporation  & Mount Sinai Hospital\isout & St Francis-Downtown & Brigham And Women's Hospital\\
		12. & North Central Bronx Hospital   & Ohio State University Hospitals & Eastern Niagara Hospital - Lockport Division & North Shore University Hospital\\
		13. &  University Hospital Of Brooklyn ( Downstate )  & Hackensack University Medical Center & Jersey Shore University Medical Center & Montefiore Medical Center\\
		14. &  Queens Hospital Center  & St Vincent Hospital \& Health Services & Mercy Catholic Medical Center- Mercy Fitzgerald & Duke University Hospital\isout\\
		15. & Montefiore Medical Center   & York Hospital\isout & Virtua Willingboro Hospital\isout &Mercy Hospital Springfield\isout\\
		16. & Phs Indian Hospital At Pine Ridge   & Houston Methodist Hospital & Easton Hospital\isout & Baylor University Medical Center\isout\\
		17. &  Brooklyn Hospital Center - Downtown Campus   & Northwestern Memorial Hospital & Cleveland Clinic\isout & Baptist Medical Center\isout \\
		18. & Regional One Health   & Memorial Hermann Hospital System & Advocate Christ Hospital \& Medical Center & Uf Health Shands Hospital\\
		19. &  Ivinson Memorial Hospital  & Methodist Hospital & Healthsource Saginaw & Houston Methodist Hospital\\
		20. &  Campbell County Memorial Hospital  & Baylor University Medical Center\isout & St Lukes Hospital Of Kansas City & University Of Michigan Health System\\
		\bottomrule
	\end{tabular}
	}
\end{table}
}

%
%
%

\end{document}